\documentclass[10pt]{article}

\usepackage[english]{babel}
\usepackage[letterpaper]{geometry}
\usepackage[utf8]{inputenc}
\usepackage{cite}
\usepackage{xspace}
\usepackage{url}
\usepackage{amsmath}
\usepackage{mathtools}
\usepackage{amssymb}
\usepackage{amsfonts}
\usepackage{bbm}
\usepackage{array}
\usepackage{mdwtab}
\usepackage{eqparbox}
\usepackage[lined,boxed,commentsnumbered,ruled]{algorithm2e}
\usepackage{tikz}
\usetikzlibrary{shapes,arrows,automata,positioning,snakes,backgrounds,petri}
\usepackage{listings}
\usepackage{caption}
\usepackage{textcomp}
\usepackage{framed}
\usepackage{subfigure}
\usepackage{makeidx}
\usepackage{framed}

\newcommand{\CL}{$\mathcal{CL}$\xspace}
\newcommand{\RCL}{$\mathcal{RCL}$\xspace}

\lstdefinelanguage{dot}{
	sensitive=true,
	comment=[l]{//},
	morecomment=[s]{/*}{*/},
	string=[b]{"},
	keywords=[1]{
		arrowhead, arrowsize, arrowtail, bgcolor, bottomlabel, center, clusterrank,
		color, comment, compound, concentrate, constraint, decorate, dir, distortion,
		fillcolor, fixedsize, fontcolor, fontname, fontpath, fontsize, group, headlabel,
		headport, headURL, height, label, labelangle, labelcolor, labeldistance,
		labelfloat, labelfontname, labelfontsize, labeljust, labelloc, layer, layers,
		lhead, ltail, margin, mclimit, minlen, nodesep, nslimit, nslimit1, ordering,
		orientation, page, pagedir, peripheries, quantum, rank, rankdir, ranksep, ratio,
		regular, remincross, rotate, samehead, sametail, samplepoints, searchsize,
		shape, shapefile, sides, size, skew, style, taillabel, tailport, tailURL,
		toplabel, URL, weight, width, z
	},
	morekeywords=[2]{
		node, edge, graph, digraph, subgraph, strict
	},
}

\lstdefinestyle{dot}{
	basicstyle=\footnotesize\ttfamily,
	numbers=left,
	numberstyle=\tiny,
	numbersep=5pt,
	frame=lines,
	breaklines=true,
	prebreak=\raisebox{0ex}[0ex][0ex]{\ensuremath{\hookleftarrow}},
	showstringspaces=false,
	upquote=true,
	tabsize=1,
}

\lstdefinelanguage{rcl}{
	keywords=[1]{ O, P, F,CONFLICT,Trace,Stacktrace},
}

\lstdefinestyle{rcl}{
	basicstyle=\normalsize,
	numbers=left,
	numberstyle=\normalsize,
	numbersep=5pt,
	frame=lines,
	breaklines=true,
	prebreak=\raisebox{0ex}[0ex][0ex]{\ensuremath{\hookleftarrow}},
	showstringspaces=false,
	upquote=true,
	tabsize=1,
}

\SetKwFunction{construirAutomato}{construirAutomato}
\SetKwFunction{buscarConflitos}{buscarConflitos}
\SetKwInOut{Input}{Input}
\SetKwInOut{Output}{Output}
\SetKwFor{Para}{para}{}{}
\SetAlCapSkip{1em}

\tikzstyle{tech} = [rectangle, draw, , node distance=3cm, text width=14em, text centered, minimum height=3em]
\tikzstyle{obj} = [rectangle, draw, node distance=3cm, text width=10em, text centered, rounded corners, minimum height=2em]  
\tikzstyle{final-obj} = [rectangle, draw, node distance=3cm, text width=14em, text centered, rounded corners, minimum height=3em]    
\tikzstyle{line} = [draw, -latex]
\tikzstyle{logic} = [draw, ellipse, node distance=3cm, text width=7em, text centered, minimum height=3em]
\tikzstyle{logic1} = [draw, ellipse, node distance=3cm, text width=3em, text centered, minimum height=2.5em]

\newcommand{\CLR}{relativized $\mathcal{CL}$\xspace}

\begin{document}
\title{An automatic tool for checking multi-party contracts}

\author{Adilson Luiz Bonifacio\thanks{Computing Department, University of Londrina, Londrina, Brazil.} \and
Wellington Aparecido Della Mura\thanks{Computing Department, University of Londrina, Londrina, Brazil}}

\date{} 

\maketitle

\begin{abstract}
Contracts play an important role in business where relationships among different parties are dictated by legal rules.
The notion of electronic contracts has emerged mostly due to technological advances and the electronic trading among companies and customers. 
Thereby new challenges have arisen to guarantee reliability among the stakeholders in electronic negotiations.
In this scenery, the automatic verification of electronic contracts appeared as the solution but as a new challenge  at the same time. 
An important task on verifying contracts is concerned of detecting conflicts in multi-party contracts.
The problem of checking contracts has been largely addressed in the literature, but we are not aware about any method and tool that deals with multi-party contracts and conflict detection using a contract language.  
This work presents an automatic checker, so-called RECALL, for finding conflicts on multi-party contracts  modeled by an extension of a contract language. 
We developed an automatic checking tool and also applied it to a a well-known case study of selling products that is characterized by multi-party aspects  of the contracts. 
We also performed some experiments in order to show the tool performance w.r.t. the size of contracts.
\end{abstract}

\section{Introduction}\label{sec:introduction}

Business relationships have become increasingly more customary among companies and customers thanks to technological advances.
In this setting new challenges have arisen on business negotiations seeing that interrelationships among the stakeholders are liable to disagreements.
To overcome potential disagreements transaction rules have been introduced to avoid conflict situations based on the notion of legal contracts. 

Contracts are composed by clauses that describe business rules in a setting with several involved parties.
Then, obligations, prohibitions, and permissions can be enforced by rules which express rights and duties over the parties of a contract. 
A contract can be, in general, classified in three groups: unilateral contracts, when only a single party assumes responsibilities; bilateral contracts, when responsibilities are upon two involved parties; and multilateral, or multi-party contracts, when several parties assume responsibilities~\cite{Angelov01b2becontract}.

Conflicts may arise in a contract when two or more parties are involved by associated rules, especially for multi-party contracts, where several parties are simultaneously related to each other. 
An ambiguous description of a contract, \emph{e.g.} given in natural language, may result in inconsistent relationships. 
Problems of this nature can be avoided or mitigated when contracts are more accurately  described by formalisms. 
A precise description of contracts guarantees more reliability in their interpretations. 
However, even making use of formalisms inconsistencies and ambiguities can be introduced in a contract specification. 
Therefore an automatic verification of contracts specified by an appropriate formalism is desirable to avoid conflicts. 

Formal verification of electronic contracts have been largely studied in the literature~\cite{Xu04amulti-party,Fenech09Automatic,Kyas08runtimemonitoring,Prisacariu09AnAction}.
Some approaches were proposed to either guarantee certain properties on the contracts or  detect undesired situations, such as deadlocks, and unnecessary or conflicting clauses.
Conflicts are characterized by contradictory clauses which, in turn,  can incur in incoherent rules of a contract. 
Hence it is desirable that a conflict may be detected and solved in a contract negotiation process before running the contract in practice.

In this work we treat a more complex class of contracts, so-called multi-party contracts which are specified by an extended contract language. 
The classical  Contract Language (CL) has been proposed by Prisacariu~\cite{Prisacariu12_a_dynamic} to comprise concepts of the relativized deontic logic~\cite{Herrestad1995Deontic} and the dynamic logic~\cite{Harel84dynamiclogic} to specify multi-party contracts. 
We then propose a conflict detection method for multi-party contracts modeled by this extended contract language, named Relativized Contract Language (\RCL)~\cite{rcl-sccc15}. 
Our method allows for more complex contracts where parties' relationships and their designations are deemed important. 
The method has been developed and named by  $\mathbf{R}$elativiz$\mathbf{E}$d $\mathbf{C}$ontr$\mathbf{A}$ct $\mathbf{L}$anguage ana$\mathbf{L}$yser (RECALL)\footnote{Available on \url{http://recallcontracts.github.io}}. 
We also model and check a real-world case study  by using the RECALL tool. 
The presented case study  aims in twofold:
first, it provides a proof of concept on the RECALL's functionalities; and secondly, it allow us to check a well-known problem in the literature. 
We further describe some practical experiments to evaluate the efficiency and effectiveness of our tool. 
We randomly generate different groups of multi-party contracts where each experiment evaluates aspects related to scalability, processing time, and resource consumption. 

The remaining of this paper is organized as follows.
We survey some works that are more closely related to the conflict detection problem for multi-party contracts in Section~\ref{sec:related_work}. 
In Section~\ref{sec:motivation} we motivate our work introducing a real case study for a sales contract. 
We give the foundations of our proposed method in Section~\ref{sec:method}. 
Section~\ref{sec:implementation} presents the RECALL tool and  its practical application over the real case study based on trading rules modeled and checked using \RCL. 
In Section~\ref{sec:experiments} some experimental results are presented to evaluated the proposal.  
Section~\ref{sec:conclusion} gives some concluding remarks and future directions.


\section{Related work} \label{sec:related_work}

Multi-party contracts are defined by agreements signed by several stakeholders in a business deal.
A contract of this nature cannot be decomposed within a set of bilateral contracts, where the arrangements are firmed by stakeholders in pairs, without loss of information~\cite{Xu04amulti-party}. 
Therefore multi-party contracts are, in fact, more complex to be modeled and verified because they require suitable formalisms to appropriately express manifold relationships and particular identifications. 
According to Fenech et al.~\cite{Fenech09Automatic}, particular responsibilities cannot be placed in a bilateral contract modeled by \CL~\cite{Prisacariu07aformal}. 

By contrast, Herrestad and Krogh~\cite{Herrestad1995Deontic} have proposed an extension of the standard deontic logic~\cite{Hilpinen2001Deontic}, so-called relativized deontic logic, to identify stakeholders in relationships. 
The relativized deontic logic personifies deontic operators, allowing for more complex scenarios when several parties are simultaneously involved in a deal. 
Unfortunately,  a formalism to precisely deal with the remaining aspects of \CL, aside from the standard deontic logic, is lacking. 

Fenech et al.~\cite{Fenech09Automatic} then proposed a conflict detection method for bilateral contracts based on \CL. 
Thus we extend their classical mechanism to support the analysis of multi-party contracts. 
First, we extend the  \CL contract representation and then we propose a  conflict detection algorithm for the extension. 

\begin{figure}[!hbpt]
	\centering
	\begin{tikzpicture}[node distance = 3cm, auto, scale=0.7,transform shape]
	\node [logic] (deologic) {{\footnotesize Deontic Logic}};     
	\node [logic1, below right of=deologic] (cl) {{\footnotesize \CL}};
	\node [logic, above right of=cl] (dynlogic) {{\footnotesize Dynamic Logic}};
	\node [tech, right=2.75cm of cl] (confDetect) {{\footnotesize \CL conflict detection}};
	\node [logic, left=2cm of cl] (rdl) {{\footnotesize Relativized Deontic Logic}};
	\node [obj, below left of=cl] (cl2) {{\footnotesize \RCL}};
	\node [final-obj, right=4cm of cl2] (detect2) {{\footnotesize Multi-party conflict detection}};
	
	\path [line] (dynlogic) -- (cl);
	\path [line] (deologic) -- (cl);
	\path [line] (deologic) -- (rdl);
	\path [line] (confDetect) -- node {using}  (cl);
	\path [line, dashed] (rdl) -- (cl2);
	\path [line, dashed] (cl)  -- (cl2);
	\path [line, dashed] (detect2) -- node {using}   (cl2);
	\path [line, dashed] (confDetect)  -- (detect2);
	\end{tikzpicture}
	\caption{Scheme of our contribution.
	 \label{fig:fluxogram}}
\end{figure}
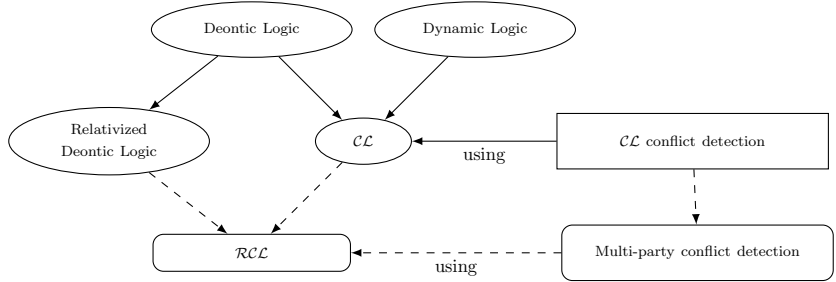

We represent our contribution and its relation with logics, languages, and techniques proposed in the literature by the scheme depicted in Figure~\ref{fig:fluxogram}. 
The rounded rectangles and dashed arrows indicate our contributions. 

\section{A Real-world Multi-party Contract}\label{sec:motivation}

In this section we start with the real-world case study that motivates our work and also supports the proof of concept in the tool testing. 
The case study  looks into a well-known business model of selling products~\cite{Daskalopulu2001}.
In the following subsections we, first, describe the electronic sales contract and in the sequel present multi-party aspects that characterize it. 

\subsection{The Sales Contract}\label{cenario}

Electronic transactions have become a fairly common practice.
Although there are several consumer protection laws, barely formulated agreements can remain on a contract in such a way that frauds and misunderstandings may still arise among the stakeholders.
Electronic commerce  become more complex when its business model contains interdependencies among the participating parties.
Interdependencies are, in fact, necessary to precisely express the complexity of a certain contract.

For instance, in a bilateral sales contract we cannot guarantee  that the seller and the buyer comply with the agreement upon the payment and the delivery of the traded product.
On the one hand, the seller cannot deliver the product to the buyer before the latter has paid it.
In this case, the seller is unprotected once the buyer in turn does not pay the product and the former is harmed.
On the other hand, if the buyer pays for the product before receiving it, then the seller might not deliver the product and the agreement comes to a violation.
In any case both are unprotected and the contract can be violated either due to the lack of payment or not receiving the goods.

Financial  agencies are then used to intermediate payments in a electronic commerce,  and thus avoiding contractual breaches.
In the present case study, financial transactions are intermediated by a bank to which payments are accomplished by a buyer but the amount is transfered to the seller only after the respective product is delivered to the buyer. 
Thus we guarantee that a buyer is not harmed since the bank can reimburse him if the product is not delivered.
In addition, internal rules and specific regulations of the involved parties can also be included in a contract.

Next we present all arrangements and rules of the sales contract. 
The stakeholders in the case study are: buyer, seller, bank and shipping company. 
A buyer buys a product from a seller, the shipping company (or carrier) is in charge to deliver the product to the buyer, and the bank intermediates all financial transactions.
Besides the agreements we need to take into account internal rules of the stakeholders.
For instance, the carrier need not to deliver an order while the shipping cost has not been paid by the seller.
In addition payments are accomplished by the bank only after it receives an appropriate notification in order to prevent frauds.
Now we fully describe the contract:
\begin{framed}
{\scriptsize
        \begin{enumerate}
            \item []\textbf{[Sales Contract]}
            \item \textbf{Buyer} performs the purchase of a product from the \textbf{Seller}.
            \item \textbf{Buyer} is obliged to pay the product to the \textbf{Bank}.
            \item \textbf{Bank} must send the notification about product's payment to  \textbf{Seller}.
            \item After \textbf{Bank} notifies the \textbf{Seller} about the payment, \textbf{Seller} is obliged to send the product by means of \textbf{Carrier} and pays the product's shipping costs to the \textbf{Bank}.
            \item \textbf{Carrier} must deliver the product to \textbf{Buyer}.
            \item After the product is delivered, \textbf{Buyer} is obliged to acknowledge the \textbf{Bank} about the product delivery, whereas  \textbf{Carrier} must notify the \textbf{Seller} that the product was delivered to \textbf{Buyer}.
            \item When the \textbf{Seller}  is notified about the product delivery, the \textbf{Seller} is obliged to notify the \textbf{Bank} allowing the payment of the shipping costs to the \textbf{Carrier}.
            \item When \textbf{Buyer} notifies the  \textbf{Bank} that the product was received,  \textbf{Bank} releases the payment amount to the \textbf{Seller}.
            \item \textbf{Bank} must pay the shipping costs to the \textbf{Carrier} after \textbf{Seller} makes the payment of the referred amount.
            \item []\textbf{[Internal Bank Rules]}
            \item \textbf{Bank} is prohibited to pay \textbf{Seller} till it has received a proper notification from \textbf{Buyer} confirming the product delivery.
            \item \textbf{Bank} is prohibited to release the payment of shipping costs for the  \textbf{Carrier} till \textbf{Seller} notifies the bank.
            \item []\textbf{[Internal Carrier Rules]}
            \item \textbf{Carrier} is prohibited to deliver the product till \textbf{Seller} has paid the shipping costs.
        \end{enumerate}}
\end{framed}

To easy the reference and to keep the notation uncluttered we define some symbols and key words for the contract. 
The stakeholders are also called by individuals and we refer them by symbols as defined in Table~\ref{tab:study_individuals}.
Similarly, actions related to the contract are given in
Table~\ref{tab:study_actions}.
\begin{table}[!ht]
\begin{minipage}{.33\textwidth}
	\centering
	\begin{tabular}{|c|c|}
		\hline
		\textbf{Individual} & \textbf{Symbol} \\ \hline
		buyer & b \\ \hline
		seller & s \\ \hline
		bank & k \\ \hline
		carrier & c \\ \hline
	\end{tabular}
	\caption{List of individuals. \label{tab:study_individuals}}
\end{minipage}
\hspace*{3ex}
\begin{minipage}{.6\textwidth}
	\centering
	\begin{tabular}{|c|l|}
		\hline
		\textbf{Action} & \textbf{Description} \\ \hline
		buyProduct & Buy a product \\ \hline
		payProduct & Pay the product  \\ \hline
		notifyProductPayment & Notify the product payment \\ \hline
		sendProduct & Send the product \\ \hline
		deliverProduct & Deliver the product \\ \hline
		notifyProductReceipt & Notify the product receipt\\ \hline
		notifyProductDelivery & Notify the product delivery\\ \hline
		payShippingCosts & Pay the shipping costs \\ \hline
		releaseShippingCosts & Release the shipping costs \\ \hline
	\end{tabular}
	\caption{List of actions \label{tab:study_actions}}
\end{minipage}
\end{table}

\subsection{Multi-party Aspects of the Contract}\label{multilateral}

Bilateral contracts are not able to establish multi-party relationships composed by more  than two individuals~\cite{Xu04_monitoring_thesys}. 
Hence a multi-party contract cannot be broken down into a set of bilateral contracts without loss of information. 
In this section we have no intention to formally prove such property, provided that other works have studied it~\cite{Xu04amulti-party,Haugen02_multy}. 
Instead we aim at showing that the sales contract is, in fact, characterized by multi-party aspects. 
Therefore we make use of Petri Net models~\cite{DBLP:conf/ifip/Petri62,Reisig:1985:PNI:3405,murata89} to specify the sales contract. 
Petri Net is a well-known formalism that can appropriately describe distributed systems which, in turn, is characterized by different components communicating to coordinate to each other in whole system.

A basic Petri Net model is defined by a collection of directed arcs connecting places and transitions. 
Places may hold tokens which represent the state or marking of a net according to its assignment by tokens to places. 
Arcs can only connect places to transitions and have capacity one by default. 
Otherwise the capacity must be explicitly marked on the arc. 
 A transition is enabled when the number of tokens in each of its input places is at least equal to the arc weight going from the place to the transition. 
Tokens of input places are moved to output places when an  enabled transition is fired according to arc weights. 
After the transition firing, a new marking of the net is obtained as the result, a state description of all places.

The Petri Net model for the sales contract is depicted in Figure~\ref{fig:petri-sales}. 
The specification basically follows the previous Sales Contract description. 
The resulting model is composed by eleven places, $p_1 \ldots p_{11}$,  nine transitions, $t_1 \ldots t_9$, and has one token marked at place $p_1$ to start the execution. 
This token at place $p_1$ enables transition $t_1$ that can be fired at any time. 
When  the buyer performs the purchase transition $t_1$ fires and produces one token at place $p_2$. 
It means that transition $t_2$ is enabled to fire, i.e. buyer can pay the product. 
When $t_2$ is fired one token is produced at places $p_3$ and $p_4$. 
It means that the buyer is waiting for the product (place $p_3$) and the payment was received by the bank (place $p_4$). 
 \tikzset{
  place/.style={
    circle,
    thick,
    draw=blue!75,
    fill=blue!20,
    minimum size=6mm
  },
  transition/.style={
    rectangle,
    thick,
    fill=yellow, 
    minimum width=8mm,
    inner ysep=2pt
    }
  }            
  \begin{figure}
  \begin{center}
  \begin{tikzpicture}[node distance=2cm,>=stealth',bend angle=45,auto]

    \node [place] (p1)   {$p_1$};
    
    \node [place] (p2)  [below = 2.4cm of p1] {$p_2$};
    
    \node [transition] (t1) [below = 1cm of p1] {$t_1$}
      edge [pre]                  (p1)
      edge [post]                (p2);


  \node [place] (p3)  [below=4cm of p2] {$p_3$};
  \node [place] (p4)  [right=5cm of p2] {$p_4$};
  
    \node [transition] (t2) [below=1.5cm of p2] {$t_2$}
      edge [pre]                  (p2)
      edge [post]                  (p3)
        edge [post]                  (p4);


    \node [place] (p5)  [below = 3cm of p4] {$p_5$};
    
    \node [transition] (t3) [below of=p4] {$t_3$}
      edge [pre]                  (p4)
      edge [post]                  (p5);


        \node [place] (p6)  [below = 3cm of p5] {$p_6$};
          \node [place] (p7)  [above left = 2cm of p6] {$p_7$};
        
       \node [transition] (t4) [below of=p5] {$t_4$}
          edge [pre]                  (p5)
          edge [post]                (p6)
          edge [post]                  (p7);          


       \node [transition] (t5) [right of=p3] {$t_5$}
          edge [pre]                  (p7)
          edge [post]                  (p3);          


  \node [place] (p8)  [below=4cm of p3] {$p_8$};
  \node [place] (p9)  [below right=2cm of p3] {$p_9$};
    
    \node [transition] (t6) [below=1.5cm of p3] {$t_6$}
      edge [pre,label=above left:$2$]                  (p3)
        edge [post]                  (p8)
        edge [post]                  (p9);


    \node [transition] (t7) [right=1cm of p9] {$t_7$}
      edge [pre]                  (p9)
        edge [post]                  (p6);


  \node [place] (p10)  [below=3cm of p6] {$p_{10}$};
    
    \node [transition] (t8) [below=1.5cm of p6] {$t_8$}
      edge [pre,label=above left:$2$]                 (p6)
        edge [post]                  (p10);


  \node [place] (p11)  [below right=2.5cm of p8] {$p_{11}$};
  
    \node [transition] (t9) [below right=1cm of p8] {$t_9$}
      edge [pre]                  (p8)
        edge [post]                  (p11);

\end{tikzpicture}
 \end{center}
\caption{Petri net for the sales contract.}
\label{fig:petri-sales}
  \end{figure}
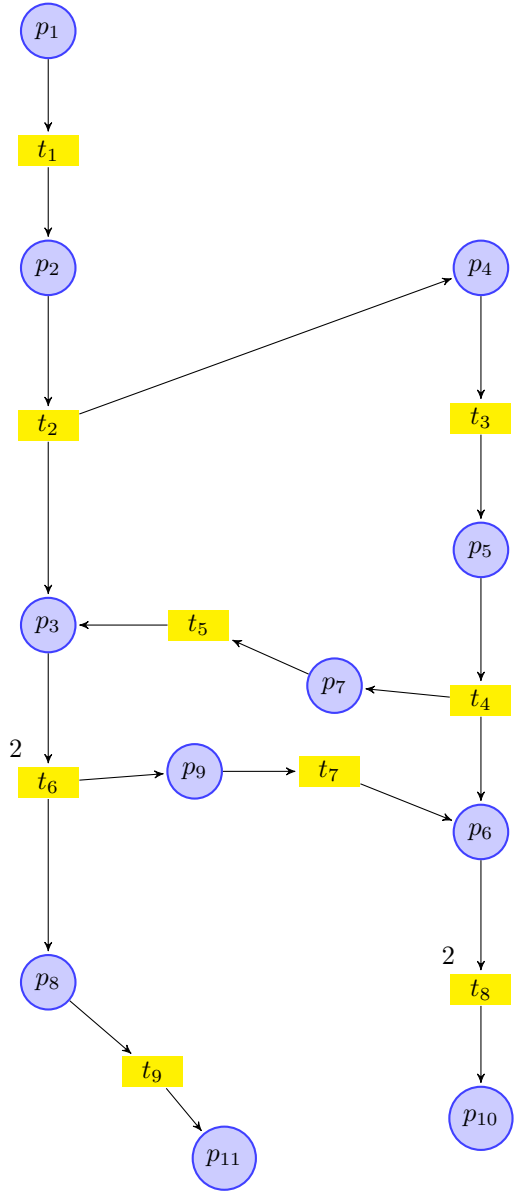

Note that the marked token at place $p_3$ is not enough to enable transition $t_6$ because its precondition (arc from $p_3$ to $t_6$) is weighted by two tokens. 
However transition $t_3$ is enabled by $p_4$ and when it is fired the bank notifies the seller about the payment, producing one token at place $p_5$. 
It represents that the seller must send the product by means of  the carrier and also must pay the shipping costs. 
When both conditions are satisfied transition $t_4$ is then fired and one token is produced in each place, $p_6$ and $p_7$. 
Place $p_6$ reflects that although the bank holds the shipping costs, the carrier cannot be payed yet (transition $t_8$), unless the product is delivered to the buyer (transition $t_7$).
But place $p_7$ denotes the product is already in tenure of the carrier and enables transition $t_5$. 
The product is effectively delivered to the buyer when transition $t_5$ is fired and another token is produced at place $p_3$. 

At this point place $p_3$ has two tokens and it enables transition $t_6$. 
After transition $t_6$ is fired one token is produced at place $p_8$, and another one is produced at place $p_9$. 
The latter says the bank is notified by the buyer as concerning the product receipt, and the former says the carrier has notified the seller about the delivery. 
Hence transition $t_7$ is enabled and after it is fired another token is produced at place $p_6$. 
Now place $p_6$ enables transition $t_8$ and when $t_8$ is fired the carrier receives the shipping costs. 
The payment is represented by the token produced at place $p_{10}$. 
On the other hand, that token produced at place $p_8$  enables transition $t_9$ which means that the bank can release the payment to the seller. 
After $t_9$ is fired one token is produced at place $p_{11}$ meaning that the seller has received the payment. 
The final marking of the net with one token at place $p_{10}$ and another token at place $p_{11}$ means the trading is already finished. 

We know that different bilateral contracts cannot share information and actions which are  preconditions cannot be assured to happen simultaneously on distinct bilateral contracts.
Thus we notice that when the sales contract is modeled by separating bilateral contracts where one individual of a subcontract is not directly related to other individual of a distinct subcontract the  expressiveness of whole business model cannot be properly captured.

It is a simple matter to see that whether we have both seller and carrier separated in two different bilateral contracts we would have a setting where  the carrier could not notify the seller about the delivery.
So one cannot guarantee that the seller would notify the bank to release  the shipping costs to the carrier which, in turn, could  be harmed in the trading.
We can observe  this scenario in the Petri Net model depicted in Figure~\ref{fig:petri-sales}.  
Had the transition $t_6$  not been fired to produce one token at place $p_9$, transition $t_7$ would not also be fired to produce the required token at place $p_6$. 
Then the transition $t_8$ which is weighted by two would not be enabled and so place $p_{10}$ would never take the required token to denote the end of the process.
Such situations render to  unstable contracts since the required relationships and preconditions cannot be properly modeled.


\section{Multi-party conflict detection method } \label{sec:method}

We extend the classical mechanism proposed by 
Fenech et al.~\cite{Fenech09Automatic} to support the analysis of multi-party contracts. 
The  conflict detection method comprises, basically, two steps: the construction of an automaton to represent the contract; 
and the conflict detection analysis that is applied on this automaton.

The conflict detection process for multi-party contracts also requires an appropriate formalism to specify agreements of rights and duties under the perspective of several stakeholders.
In this case, business rules rely on specific parties of a contract, in contrast to bilateral relationships where rights and duties are globally assumed in the whole contract. 
The extension proposed in this work comprises on: 
extending the \CL syntax to support relativization; 
defining a semantics for \CL operators according to new syntax; 
modifying  the automaton construction based on the new semantics; and 
adapting the conflict detection algorithm regarding the new trace semantics. 
Next subsections detail each step of our method and also give part of the case study of a sales contract as a running example. 
We make it clear that our work has a more practical leaning and it is not our intention here to formally prove the correctness of the proposal.


\subsection{Extension of~\CL syntax} \label{subsec:newSyntax}

Revitalizations upon deontic operators  were proposed by 
Herrestad and Krogh~\cite{Herrestad1995Deontic}. 
Particular individuals can  be associated to deontic operators 
to allow individual identification when actions are performed in a contract. 
Beyond the global operators, as defined in the standard deontic logic, 
relativized operators can explicitly specify senders, receivers, or both on the operators. 
The classical \CL is defined over the standard deontic logic and over the dynamic logic, 
but no relativization is allowed by the language. 

A multi-party contract can be defined by a set of clauses, which in turn, are composed by obligations, permissions, prohibitions, and dynamic modalities over actions. 
Formally, let $\mathcal{C}$ be a contract, where $\mathcal{I}$ is the set of individuals (parties) of the contract. 
The set of deontic modal operators is $\mathcal{D} = \{O, P, F\}$, where $O$ is an obligation, $P$) is a permission and $F$ is a prohibition. 
The set $\mathcal{R} = \{g, i, i\curvearrowright j \mid i,j \in \mathcal{I}\}$ defines all possible relativizations, and $\mathcal{A_B}$ is the set of basic actions, where $\alpha \in \mathcal{A_B}$ is an action associated to the operator $d$, with $d \in \mathcal{D}$. 
For instance, $_gd(\alpha)$ indicates that the action $\alpha$ must be performed by all individuals of $\mathcal{C}$ according to the deontic operator $d$. 
On the other hand,  $_id(\alpha)$ indicates that the action $\alpha$, associated to $d$, must be performed by $i \in \mathcal{I}$. 
Finally, $_{i\curvearrowright j}d(\alpha)$ says that the action $\alpha$ associated to operator $d$ must be performed by $i$ and received by $j$. 
For the sake of simplicity, we omit the symbol $g$, which represents a global operator. 
Considering dynamic operators, $_g[\alpha]\mathcal{C}$ indicates that after action $\alpha$ is performed, a contract $\mathcal{C}$ takes place. 
The relativized dynamic operator $_i[\alpha]\mathcal{C}$ denotes that $\mathcal{C}$ is valid if the individual $i$ performs $\alpha$. 
Lastly, $_{i \curvearrowright j}[\alpha]\mathcal{C}$ indicates that $\mathcal{C}$ takes place if $i$ performs $\alpha$ to $j$.

The \RCL syntax is presented in Figure~\ref{eq:gramCLR}.
\begin{figure}[!htpb]
\begin{align*}
&\mathcal{C}::=\mathcal{C}_O \mid \mathcal{C}_P \mid \mathcal{C}_F \mid \mathcal{C} \wedge \mathcal{C} \mid \mathcal{C}_D\mid \top \mid \perp\\
&\mathcal{C}_O::= O_\mathcal{C}(\alpha) \mid \;_iO_\mathcal{C}(\alpha) \mid \;_{i \curvearrowright  j}O_\mathcal{C}(\alpha) \mid \mathcal{C}_O \oplus \mathcal{C}_O \\
&\mathcal{C}_P::= P(\alpha) \mid \;_iP(\alpha) \mid \;_{i \curvearrowright  j}P(\alpha) \mid \mathcal{C}_P \oplus \mathcal{C}_P\\
&\mathcal{C}_F::=F_\mathcal{C}(\alpha) \mid \;_iF_\mathcal{C}(\alpha) \mid \;_{i \curvearrowright  j}F_\mathcal{C}(\alpha) \mid \mathcal{C}_F \vee \mathcal{C}_D \mathcal{C}_F \\
&\mathcal{C}_D ::=[\beta]\mathcal{C} \mid \;_i[\beta]\mathcal{C} \mid \;_{i \curvearrowright  j}[\beta]\mathcal{C} \\
&\alpha::= 0 \mid 1 \mid a \mid \alpha \times \alpha \mid \alpha \cdot \alpha \mid \alpha + \alpha\\
&\beta::=  0 \mid 1 \mid a \mid \beta \times \beta \mid \beta \cdot \beta \mid \beta + \beta \mid \overline{\beta}  \mid \beta^*
\end{align*}
\caption{Relativized \CL Grammar \label{eq:gramCLR}}
\end{figure}
A contract $\mathcal{C}$ can then be derived following this grammar according to the conventional definitions. 
Penalty mechanisms, denoted by a subscript $\mathcal{C}$, can also be defined on operators of obligation and prohibition when violations occur. 

We also note that actions can be composed by different operators: 
choice between two actions denoted by $+$, concurrency between actions specified by $\times$,  priority on actions given by $\cdot$, as well as the special actions $\mathbf{0}$ and $\mathbf{1}$ to represent, respectively, the contract violation and the execution of any action. 
A dynamic operator allows the iteration operator $\ast$ over an action, and the negation of an action $\alpha$, which means that any other action can be performed instead of  $\alpha$.

Decompositions and equivalences obtained from the grammar extension, likewise in \CL, define the semantics. 
Figure~\ref{fig:decomp} presents some decompositions over \RCL, and the 
equivalences follow as proposed by Meyer~\cite{Meyer1987ADifferentApproach}. 
\begin{figure}[!htpb]
\begin{align*}
&\;_iO_\mathcal{C}(\alpha) \iff \;_i[\overline{\alpha}]\mathcal{C}\\
&\;_iF_\mathcal{C}(\alpha) \iff \;_i[\alpha]\mathcal{C}\\
&\;_iO_\mathcal{C}(\alpha\times \beta) \iff_iO_\mathcal{C}(\alpha) \wedge \;_iO_\mathcal{C}(\beta)\\
&\;_iO_\mathcal{C}(\alpha\cdot \beta) \iff_iO_\mathcal{C}(\alpha) \wedge \;_i[\alpha]\;_iO_\mathcal{C}(\beta)\\
&\;_iO_\mathcal{C}(\alpha+ \beta) \iff (\;_iO_\mathcal{C}(\alpha) \wedge \;_iO_\mathcal{C}(\beta)) \oplus \;_iO_\mathcal{C}(\alpha) \oplus \;_iO_\mathcal{C}(\beta)\\
&\;_iF_\mathcal{C}(\alpha\times\beta) \iff \;_iF_\mathcal{C}(\alpha) \wedge \;_iF_\mathcal{C}(\beta)\\
&\;_iF_\mathcal{C}(\alpha\cdot\beta) \iff \;_iF_\mathcal{C}(\alpha) \vee \;_i[\alpha]\;_iF_\mathcal{C}(\beta)\\
&\;_iF_\mathcal{C}(\alpha+\beta) \iff \;_iF_\mathcal{C}(\alpha) \wedge \;_iF_\mathcal{C}(\beta)\\
&\;_iP(\alpha\times \beta) \iff_iP(\alpha) \wedge \;_iP(\beta)\\
&\;_iP(\alpha\cdot \beta) \iff_iP(\alpha) \wedge \;_i[\alpha]\;_iP(\beta)\\
&\;_iP(\alpha+\beta) \iff \;_iP(\alpha) \wedge \;_iP(\beta)\\
&\;_i[\alpha\times \beta]\mathcal{C}  \iff \;_i[\alpha]\mathcal{C} \wedge \;_i[\beta]\mathcal{C} \\
&\;_i[\alpha\cdot \beta]\mathcal{C} \iff \;_i[\alpha]\;_i[\beta]\mathcal{C}\\
&\;_i[\alpha+\beta]\mathcal{C} \iff \;_i[\alpha]\mathcal{C} \wedge \;_i[\beta]\mathcal{C} 
\end{align*}
\caption{Decompositions of \RCL \label{fig:decomp}}
\end{figure}

For the sake of illustration  we completely specify the sales contract according to the \RCL Syntax as depicted in Figure~\ref{fig:src_studycase}. 
For instance, see the clause described at line (10)  now specified at line (19) in \RCL.
\begin{figure}[!htpb]
	\centering
	\begin{tabular}{|c|}
		\hline
\begin{lstlisting}[language=rcl]
{b,s}[buyProduct](
 {b,k}O(payProduct)^
 {b,k}[payProduct](
  {k,s}O(notifyProductPayment)^
  {k,s}[notifyProductPayment](
   {s,c}O(sendProduct)^
   {s,k}O(payShippingCosts)^
   {s,k}[payShippingCosts](
    {s,c}[sendProduct](
     {c,b}O(deliverProduct) ^
     {c,b}[deliverProduct](
      {b, k}O(notifyProductReceipt) ^
      {c, s}O(notifyProductDelivery) ^
      {b,k}[notifyProductReceipt]({k,c}O(payProduct)) ^
      {c, s}[notifyProductDelivery](
       {s, k}O(liberateShippingCosts) ^
       {s, k}[liberateShippingCosts]({k,c}O(payShippingCosts))
)))))));
{b,k}[(!notifyDelivery)*]({k,s}F(payProduct));
{s,k}[(!liberateShippingCosts)*]({k,c}F(payShippingCosts));
{s,c}[(!payShippingCosts)*]({c,b}F(deliverProduct));
\end{lstlisting}\\
		\hline
	\end{tabular}
	\caption{The \RCL sales contract specification}
	\label{fig:src_studycase}
\end{figure}


\subsection{\RCL semantics}\label{subsec:newsemantics}

We extend the trace semantics proposed by Fenech et al.~\cite{Fenech09Automatic} now upon the notion of relativization given by Herrestad and Krogh~\cite{Herrestad1995Deontic}. 
The trace semantics on \RCL  is now defined by sequences of actions, which can be relativized,  that run in a contract. 
A trace is composed by a pair: a action trace and a deontic trace. 
The action trace is composed by relativized actions putting together information upon the action associated to an operator, as well as the involved  parties. 
The deontic traces are also preserved over the deontic operators with relativizations. 
Hence the extension allow us to deal with relativized operators of \CLR syntax.

Formally speaking, a relativized action is given by a tuple $a_r = \langle\kappa,\chi,\lambda\rangle$, where $\chi \in \mathcal{A_B}$ is a basic action and $\kappa,\lambda \in \mathcal{I}$ are individuals, a sender and a receiver, respectively, associated to $\chi$. 
The set of relativized actions is then obtained by $\mathcal{A}_r = \{\mathcal{I} \times \mathcal{A_B} \times \mathcal{I}\}$, combining basic actions and individuals.
In addition, the set of concurrent relativized actions is obtained by the power set $\mathcal{A}^2_r = 2^{\mathcal{A}_r}$, where concurrent relativized actions can be combined to each other.
An action  trace is, formally, defined by $\sigma: \mathbb{N} \to \mathcal{A}^2_r$, where $\sigma$ gives the concurrent actions at position $i\in \mathbb{N}$ on the trace. 
Let $\sigma = \alpha_0, \alpha_1, \dots$ be a trace with $\alpha_i \in \mathcal{A}^2_r$, $i\geq 0$, we get $\sigma(i)=\alpha_i$. 
The length of a trace is $|\sigma|$ and the empty trace is denoted by $\varepsilon$. 
A sub-trace is written $\sigma(i..j)$, where $i$ is the initial position and $j$ is the final position on the trace. 
An infinite trace $\sigma(i...)$ denotes a sub-trace starting at position $i$. 
The concatenation of two traces $\sigma'$ and $\sigma''$ follows the standard definition, denoted by $\sigma'\sigma''$.
The union operation over two deontic traces $\sigma_d \cup \sigma_d'$ is defined by $\sigma_d(0) \cup \sigma_d'(0);\sigma_d(1) \cup \sigma_d'(1); \dots;\sigma_d(n) \cup \sigma_d'(n)$ with $\mid\sigma_d\mid = \mid\sigma_d'\mid$. 
The set of deontic modalities upon actions is given by $\mathcal{M} = \{\;_rd_\alpha \mid r\in \mathcal{R}, d \in \mathcal{D}, \alpha \in \mathcal{A_B} \}$. 
Thus $\sigma_d: \mathbb{N} \to 2^\mathcal{M}$ denotes the deontic, where each position of $\sigma_d$ gives a set of representation combinations of deontic modalities, with $\sigma_d(i) \in 2^\mathcal{M}$ for all $i \in \mathbb{N}$. 
Representations of  $2^\mathcal{M}$ distinguish conjunctions and disjunctions over the deontic modalities. 

To illustrate the definitions and notation, let 
$\mathcal{C} = \;_{i \curvearrowright j}O(a)\wedge\;_{i \curvearrowright j}O(b)\wedge\;_{i \curvearrowright j}F(b)$ and $\mathcal{C'} = \;_{i \curvearrowright j}O(a+b)\wedge\;_{i \curvearrowright j}F(b)$ be contracts. 
The deontic traces are  $\sigma_d = \langle \{_{i \curvearrowright j}O_a\},\{_{i \curvearrowright j}O_b\},\{_{i \curvearrowright j}F_b \} \rangle$ and $\sigma_d' = \langle \{_{i \curvearrowright j}O_a,\;_{i \curvearrowright j}O_b\},\{_{i \curvearrowright j}F_b \} \rangle$, respectively. 
In the former, we have a conflict due to the obligation and the prohibition over the action $b$. 
In contrast, in the latter contract any conflict can be found since the choice operator into the obligation over actions $a$ and $b$ does not violate the prohibition over $b$. 

The new trace semantics also deals with global deontic modalities, where a  concurrent relativized action denotes that the action associated to the modality must be performed by all individuals of the contract. 
For instance, $O(\alpha)$ generates the action trace $\sigma(0) \subseteq \{\langle x ,\alpha, y\rangle \mid y \in \mathcal{I}, \forall x \in \mathcal{I}\}$. 
If the modality is relativized, i.e. $_{i}O(\alpha)$, we get the trace $\sigma(0) \subseteq \{\langle i,\alpha, x\rangle \mid x \in \mathcal{I}\}$. 
Similarly, a directed modality $_{i\curvearrowright j}O(\alpha)$ produces the trace $\sigma(0) \subseteq \{\langle i,\alpha,j\rangle \mid i,j \in \mathcal{I}\}$.

We also redefine the satisfaction relation $\sigma, \sigma_d \models \mathcal{C}$ on the semantics of operators to determine whether the action trace $\sigma$ and the deontic trace $\sigma_d$ satisfy the contract $\mathcal{C}$. 
The \RCL  trace semantics is partially described in Figure~\ref{fig:semantics}, where $\mathcal{C}$ is a clause (or contract), $\alpha \in \mathcal{A}_r$ is a relativized action and $i,j \in \mathcal{I}$ are individuals. 
Formulas comprise actions that are composed by the operators of choice, concurrency and sequence (See Subsection~\ref{subsec:newSyntax}).
 \begin{figure}[!htpb]
 		\begin{align*}
 		(1) & \sigma,\sigma_d \not\models \mathcal{C}  \mbox{ if } \mid\sigma\mid\;\neq\;\mid\sigma_d\mid\\
 		(2) & \sigma,\sigma_d \models \mathcal{C}  \mbox{ if } \mid\sigma\mid = 0 \mbox{ and } \mid\sigma_d\mid = 0\\
 		(3) & \sigma,\sigma_d \models \mathcal{C}_1 \wedge \mathcal{C}_2 \mbox{ if } \sigma,\sigma_d' \models \mathcal{C}_1 \mbox{ and } \sigma,\sigma_d'' \models \mathcal{C}_2 \mbox{ and } \sigma_d = \sigma_d' \cup \sigma_d''\\
 		(4) & \sigma,\sigma_d \models O_\mathcal{C}(\alpha) \mbox{ if } \;O_{\alpha} \in \sigma_d(0) \mbox{ and }  
 		(\forall i \in \mathcal{I}, \exists \varphi \in \sigma(0), x \in \mathcal{I} \mid \\& 
 		\varphi = \langle i, \alpha, x\rangle \mbox{ and }  \sigma(1\dots),\sigma_d(1\dots) \models \top) \mbox{ or } (\sigma(1\dots),\sigma_d(1\dots) \models \mathcal{C})
 		\\
 		(5) & \sigma,\sigma_d \models \;_iO_\mathcal{C}(\alpha) \mbox{ if } \;_iO_{\alpha}\in\sigma_d(0) \mbox{ and } (\exists \varphi \in \sigma(0),  x\in \mathcal{I} \mid \varphi = \langle i,\alpha,x\rangle \\&   \mbox{ and } \sigma(1\dots),\sigma_d(1\dots) \models \top) \mbox{ or } (\sigma(1\dots),\sigma_d(1\dots) \models \mathcal{C})
 		\\
 		(6) & \sigma,\sigma_d \models \;_{i\curvearrowright j}O_\mathcal{C}(\alpha) \mbox{ if } \;_{i\curvearrowright j}O_{\alpha}\in\sigma_d(0) \mbox{ and } (\exists \varphi \in \sigma(0) \mid \varphi = \langle i, \alpha, j \rangle   \\& \mbox{ and } \sigma(1\dots),\sigma_d(1\dots) \models \top) \mbox{ or } (\sigma(1\dots),\sigma_d(1\dots) \models \mathcal{C})
 		\\
 		(7) & \sigma,\sigma_d \models [\alpha]\mathcal{C} \mbox{ if } (\forall i \in \mathcal{I}, \exists \varphi \in \sigma(0), x \in \mathcal{I} \mid \varphi = \langle i, \alpha, x\rangle  \mbox{ and } \\&
 		\sigma(1\dots),\sigma_d(1\dots) \models \mathcal{C}) \mbox{ or } (\forall i \in \mathcal{I}, x \in \mathcal{I}, \nexists \varphi \in \sigma(0)\mid \varphi = \langle i, \alpha, x\rangle)
 		\\
 		(8) & \sigma,\sigma_d \models \;_i[\alpha]\mathcal{C}\mbox{ if }(\exists \varphi \in \sigma(0), x \in \mathcal{I} \mid \varphi = \langle i, \alpha, x\rangle  \mbox{ and } \\&
 		\sigma(1\dots),\sigma_d(1\dots) \models \mathcal{C}) \mbox{ or } (x \in \mathcal{I}, \nexists \varphi \in \sigma(0)\mid \varphi = \langle i, \alpha, x\rangle)
 		\\
 		(9) & \sigma,\sigma_d \models \;_{i \curvearrowright} j[\alpha]\mathcal{C} \mbox{ if }(\exists \varphi \in \sigma(0) \mid \varphi = \langle i, \alpha, j\rangle \mbox{ and } \\&
 		\sigma(1\dots),\sigma_d(1\dots) \models \mathcal{C}) \mbox{ or } (\nexists \varphi \in \sigma(0)\mid \varphi = \langle i, \alpha, j \rangle)
 		\end{align*}
 	\caption{\RCL semantics \label{fig:semantics}}
 \end{figure}

The satisfaction relation follows the rules of the classical semantics
However, specific individuals need to be checked on the trace when operators are relativized with a sender, a receiver, or both,  to satisfy the formula. 
In the following example, $\sigma,\sigma_d \models _{i\curvearrowright j}O(\alpha)$, at least one relativized action in $\sigma(0)$ must be the action $\alpha$ which, in turn, must be performed by $i$ to another individual $j$. 
Formally, we have $\exists \varphi \in \sigma(0) \mid \varphi = \langle i, \alpha,j\rangle$. 
A dynamic modality, $\sigma,\sigma_d \models _{i\curvearrowright j}[\alpha]\mathcal{C}$, does not introduce deontic information on deontic traces. 
Therefore one considers only the action trace in the satisfaction verification. 
The remaining modalities of \RCL are obtained according to the standard derivation as described in the original \CL semantics~\cite{Fenech09Automatic}. 

\subsection{Automaton construction } \label{subsec:automaton}

To proceed with the process of detecting conflicts on a contract an automaton thats represents it must be constructed. 
We adapt the classical algorithm~\cite{Fenech09Automatic}  to detect conflicts on multi-party contracts described by \RCL. 
New deontic operators are taken into account by the algorithm and new criteria with relativizations are considered on detecting conflicts.

We formally define an automaton $\mathcal{A}$, representing a contract $\mathcal{C}$, by $\mathcal{A}(\mathcal{C})= \langle S, \mathcal{A}^2_r, \mathcal{M}, \mathcal{I}, s_0, T, V, l, \delta \rangle$, where $S$ is the set of states, $\mathcal{A}^2_r$ is the set of concurrent relativized actions, $\mathcal{M}$ is the set of deontic labels, $\mathcal{I}$ is the set of individuals, $s_0$ is the initial state, $T \subseteq S \times \mathcal{A}^2_r \times S$ is the labeled transition relation, $V$ is the violation state, $l : S \to \mathcal{C}$ is a labeling function of states with contract decompositions, and $\delta: S \to 2^{\mathcal{M}}$ is a labeling function of states with deontic information from  the decompositions. 
A sequence of concurrent relativized actions from $\mathcal{A}^2_r$ that defines the action trace is a string of the accepted language by the generated automaton. 

The construction is given by function $f : \mathcal{C} \times \mathcal{A}^2_r \to \mathcal{C}$ and it is described by Algorithm~\ref{alg:constructAutomaton}. 
Function $f$ decomposes a given contract according to the relativized actions to be performed on it. 
We illustrate a decomposition over a relativized dynamic operator and a  relativized action $\varphi \in \mathcal{A}^2_r$. 
We get $f(_{i}[\alpha]\mathcal{C}, \varphi)= \mathcal{C}$ when  $\exists a_r \in \varphi \mid a_r = \langle i,\alpha,x \rangle$ and $i,x\in \mathcal{I}$. 
Otherwise, $f(_{i}[\alpha]\mathcal{C}, \varphi)=  \top$. 	
Note that the complete definition of $f$ also considers global and directed dynamic operators, deontic operators of obligation \cite{RT-DC-15-01}, prohibition and permission as well as compound actions of dynamic operators according to the syntax and the decomposition rules of Subsection~\ref{subsec:newSyntax}.
\begin{algorithm}[!htpb] 
\SetKwFunction{constructAutomaton}{constructAutomaton}
\SetKwFunction{searchConflicts}{searchConflicts}
\SetKwInOut{Input}{input}\SetKwInOut{Output}{output}
\Input{State $s \in S$ of $\mathcal{A(C)}$} 
\Output{Automaton $\mathcal{A(C)}$}
\Begin{
	
	\uIf{\searchConflicts(s)}{
		conflict found in state $s$\;
	}\uElseIf{$l(s) = \top$}{
		$T \gets T \cup (s,\top,s)$\;
	}\uElseIf{$l(s) = \bot$}{
		$V \gets s$\;
		$T \gets T \cup (V,\perp,V)$\;
	}\Else{
		\For{$\alpha \in \mathcal{A}^2_r$}{
			$\mathcal{C}'\gets f(l(s),\alpha)$\;
			\uIf{$\exists s' \in S \mid l(s') = \mathcal{C}'$}{
				$T \gets T \cup (s,\alpha,s')$\;
			}\Else{
			$S \gets S \cup s'$\;
			$l(s') \gets \mathcal{C}'$\;			
			$T \gets T \cup (s,\alpha,s')$\;
			$\delta(s') \gets f_d(\mathcal{C}')$\;
			\constructAutomaton($s'$)\;
			}
		}
	}
	\Return $\mathcal{A(C)}$;
}
\caption{Contract automaton construction \label{alg:constructAutomaton}}
\end{algorithm}

Algorithm~\ref{alg:constructAutomaton} also makes use of an auxiliary function for deontic labeling. 
The function $f_d: \mathcal{C} \to 2^{\mathcal{M}}$ labels every state of the  automaton after a decomposition with deontic information according to the deontic trace $\sigma_d$. 
Given a contract, $f_d$ then returns a subset of $2^\mathcal{M}$ that provides the decomposed contract at a state with the deontic operators and their respective actions. 
An empty set is returned by the function when only dynamic operators and no deontic information are presented on the contract. 
Deontic information obtained by $f_d$ are at that time associated to states that represent the decompositions obtained by function $\delta$. 
The whole process stops either when a conflict is found or if whole contract were decomposed into atomic clauses. 
An atomic clause of a contract is a clause that contains only basic actions associated to deontic and dynamic operators. 
At this time, the atomic clauses are evaluated to satisfaction, violation or conflict. 

\begin{figure}[!htpb]
    \centering
    \includegraphics[width=\linewidth,height=5cm]{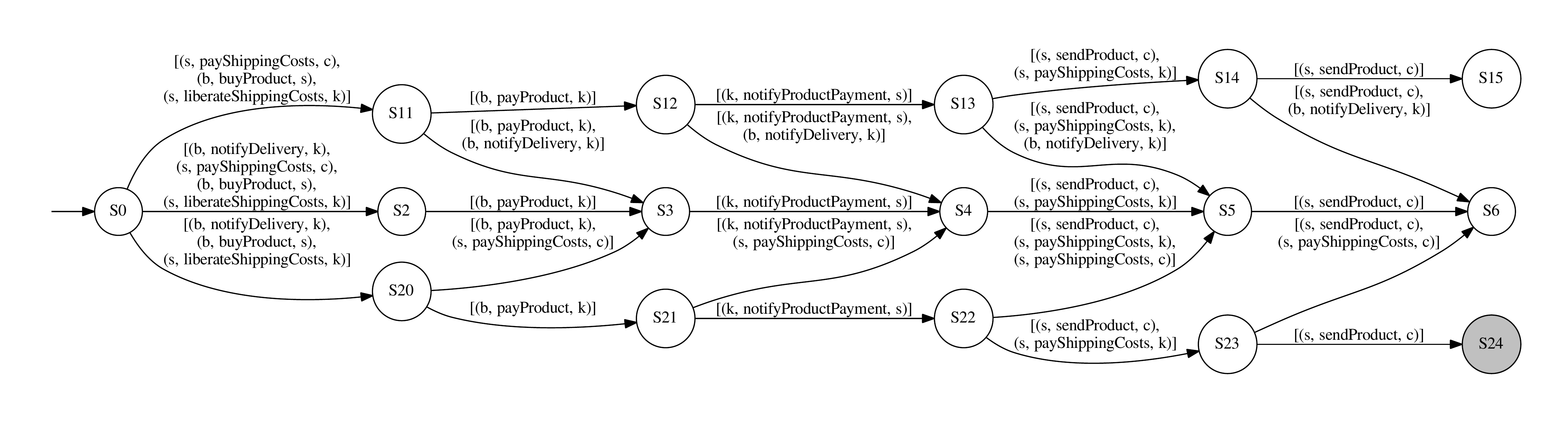}
    \caption{Partial generated automaton of the contract}
    \label{fig:study-conflicting-automaton}
\end{figure}

Note that the trace semantics is obtained according to the construction of the automaton. 
States and transitions of the automaton are constructed whereas their sequences give the semantics of traces that represents the contract as seen at line (4) in Figure~\ref{fig:src_output}. 
The corresponding automaton of the sales contract is partially depicted in Figure~\ref{fig:study-conflicting-automaton}. 
Each state of the automaton contains its respective decomposition of the contract i.e., a \RCL  formula, and an ongoing transition to it represents the set of relativized actions that was applied to obtain this formula.
Unfilled states represent intermediary decompositions obtained after evaluating their respective clauses.
A special state represents a contract violation when the evaluated formula returns false.
Another special state indicates the satisfaction of the contract, where the evaluated formula is reduced to true.
Finally the last special state that is highlighted in gray at Figure~\ref{fig:study-conflicting-automaton} indicates a conflicting decomposition as seen at line (3) on Figure~\ref{fig:src_output}.

\subsection{Conflict detection}\label{subsec:detection}

Next the conflict detection process is applied according as the algorithm constructs the automaton. 
The corresponding contract to a decomposition at a generated state is then checked in order to find conflicts.
Sceneries of conflicts treated by the algorithm are characterized by the occurrence of: 
(1) deontic operators of obligation and prohibition on the same action; 
(2) deontic operators of prohibition and permission on the same action; 
(3) deontic operators of obligation on pre-defined  conflicting actions; and 
(4) deontic operators of permission and obligation on pre-defined conflicting actions. 
A pre-defined conflicting action is, in fact, characterized when actions are performed by the same individual specified in a relativization.

In the classical approach a conflict is characterized by deontic operators occurring over the same action. 
However, a conflict does not accomplish  when  relativized deontic operators are associated to the same action but they are performed by distinct individuals. 
For instance, in the contract $\;_iO(\alpha) \wedge\;_jF(\alpha)$ where the action $\alpha$ is obliged and prohibited, simultaneously, a conflict is not characterized since different individuals are associated to the operators. 
Similarly, an action performed under operators of permission and prohibition at the same time, and fired by distinct individuals, will not induce a conflict.

We note that global deontic operators 
are always in conflict with  relativized prohibition operators. 
A conflict can then be characterized by a global obligation together with a relativized prohibition associated to the same action in a conjunction, e.g.,  $\forall x \in \mathcal{I}, \;_xO(\alpha) \wedge\; _iF(\alpha)$ or simply $O(\alpha) \wedge\; _iF(\alpha)$. 
Similarly to permission and prohibition operators when at least one of them is a global modality.

Thus the detection mechanism has been also modified to deal with for pre-defined actions in the presence of relativizations. 
The conflict relation is now defined by  $\# \subseteq \mathcal{A}_r \times  \mathcal{A}_r$ instead of the classical $\# \subseteq \mathcal{A_B} \times \mathcal{A_B}$, where $\alpha\#\beta$ denotes that actions $\alpha$ and $\beta$ can not be performed concurrently. 
Thus pre-defined conflicts among relativized actions can be described in a contract.
Relativized actions are in conflict if they are performed by the same individual. 
We can defined a conflict relation in two forms to prevent both actions being performed by a single individual: 
by a \emph{global conflict relation} when conflicting actions cannot be concurrently performed  whatever be the senders; and 
by a \emph{relativized conflict relation} when conflicting actions cannot be performed by the same individual.
The global conflict relation is denoted by $\#_g \subseteq \mathcal{A_B} \times \mathcal{A_B}$ and follows as the original relation for \CL. 
The relation  $\alpha\#_g\;\beta$, $\alpha, \beta \in \mathcal{A_B}$, denotes that actions $\alpha$ and $\beta$ cannot happen at the same time whatever be the senders. 
Similarly, the relativized conflict relation is denoted by $\#_r \subseteq \mathcal{A_B} \times \mathcal{A_B}$.
The relation $\alpha,\beta \in \mathcal{A_B}$,  $\alpha\#_r\;\beta$ denotes that the actions cannot occur simultaneously when performed by the same individual. 
Thus in a relativized conflict relation if the actions are performed by distinct individuals then a conflict does not occur.

We then proceed by redefining the conflict detection procedure described by Algorithm~\ref{alg:searchConflicts}. 
An evaluation on a state $s \in S$ of the automaton $\mathcal{A(C)}$ considers the deontic information obtained by the function $\delta(s)$. 
For every set of deontic labels $\mathcal{D} \in \delta(s)$, a state $s$ of $\mathcal{A(C)}$ has a conflict if there exists an element $d\in D$ which is in conflict to an element $d'$ of $\mathcal{D}' \in \delta(s)-\mathcal{D}$.
\begin{algorithm}[!htpb] 
{\tiny 
	\SetKwInOut{Input}{input}\SetKwInOut{Output}{output}
	\Input{A state $s$ of $\mathcal{A(C)}$}
	\Output{A conflict}
	\Begin{
		\For{$\mathcal{D}\in\delta(s)$}{
			\For{$\mathcal{D}' \in (\delta(s)-\{\mathcal{D}\})$}{
				\If{$\exists d \in \mathcal{D} \mid f_\#(d)\cap \mathcal{D}' \neq \emptyset$}{
					\Return Conflict between $d$ and $f_\#(d)\cap \mathcal{D}'$\;
				}
			}
		}
	\Return no conflict detected\;
}}
\caption{Conflict detection in \RCL \label{alg:searchConflicts}}
\end{algorithm}
The set of conflicting  deontic operators with respect to  another operator is obtained by the function $f_\#:\mathcal{M} \to 2^\mathcal{M}$~\cite{RT-DC-15-01}. 
Given a deontic operator $d$, $f_\#$ returns operators that are in conflict to $d$. 
For each $d \in \mathcal{D}$, the algorithm searches for an element of $\mathcal{D'}$ in the set of operators obtained by $f_\#(d)$. 

In our case study, the conflict detection process has found that the internal rule of the carrier conflicts with the remaining contract as described at line (20) of  Figure~\ref{fig:src_studycase}.
This internal rule states that the carrier only sends the product after the payment of the shipping costs.
On the other hand, in the original contract, one expects that the product will be delivered by the carrier to the buyer before the payment of shipping costs is released by the bank.
Therefore an incompatibility happens on the internal rule of the carrier, where the payment of the seller related to shipping costs was expected by the carrier before sending the product to the buyer.
The conflict emerges due to the internal rule of the carrier that waits the payment of shipping  costs by the seller, instead of the bank, to deliver the product.

The output analysis obtained by the practical tool (See Section~\ref{sec:implementation}) reveals such conflict as given in Figure~\ref{fig:src_output}.
\begin{figure}[!htpb]
	\centering
	\begin{tabular}{|c|}
		\hline
\begin{lstlisting}[language=rcl]
[CONFLICT] A conflict was found in the analyzed contract.
Conflict found in state (s24)
Conflict: F(c,deliverProduct,b) conflicts with [O(c,deliverProduct,b)]
Trace: (s24)<-T19-(s23)<-T17-(s22)<-T13-(s21)<-T12-(s20)<-T11-(s0)
\end{lstlisting}\\
		\hline
	\end{tabular}
	\caption{The RECALL output analysis}
	\label{fig:src_output}
\end{figure}
At the at line (1) of Figure~\ref{fig:src_output} we see the message declaring that a conflict was found in the contract.
Line (2) says in which state the conflict was found and Line (3) gives the precise information about the conflict that was detected, i.e. an atomic and subcontract.
We notice that the first clause specifies that the carrier is forbidden to deliver the product to the buyer.
At the same time, in the second clause, it is required that the carrier is obligated to deliver the product to the buyer.
Clearly, both clauses cannot be simultaneously satisfied. 

An automaton trace is shown at line (4) as a sequence of states and transitions that leads the contract to a conflicting state.
For instance, the entire contract was processed and decomposed at the initial state ($s_0$). 
According to the action on transition ($T_{11}$) the state ($s_{20}$) is obtained.
Similarly, successive decompositions are held until reaching the conflicting state ($s_{24}$).
More details are also provided by the tool such as decompositions associated to each state of the automaton and  actions that were performed over the transitions.
Figure~\ref{fig:src_conflict} shows the decomposition of the contract at state $s_{23}$ into state $s_{24}$ by transition $T_{19}$, when the seller sends the product to the carrier. 
\begin{figure}[!htb]
	\centering
	\begin{tabular}{|c|}
		\hline
\begin{lstlisting}[language=rcl]
(s24) - {s,c}[!payShippingCosts*]({c,b}F(deliverProduct)_/F/_) AND 
{c,b}[deliverProduct]({c,s}[notifyProductDelivery](
{s,k}[releaseShippingCosts]({k,c}O(payShippingCosts)_/F/_) AND 
{s,k}O(releaseShippingCosts)_/F/_) AND {b,k}[notifyProductReceipt]
({k,c}O(payProduct)_/F/_) AND {c,s}O(notifyProductDelivery)_/F/_ AND 
{b,k}O(notifyProductReceipt)_/F/_) AND {c,b}O(deliverProduct)_/F/_
<T19> - [(s, sendProduct, c), (s, payShippingCosts, k)]
(s23) - {s,c}[!payShippingCosts*]({c,b}F(deliverProduct)_/F/_) AND 
{s,c}[sendProduct]({c,b}[deliverProduct]({c,s}[notifyProductDelivery](
{s,k}[releaseShippingCosts]({k,c}O(payShippingCosts)_/F/_) AND 
{s,k}O(releaseShippingCosts)_/F/_) AND {b,k}[notifyProductReceipt]
({k,c}O(payProduct)_/F/_) AND {c,s}O(notifyProductDelivery)_/F/_ AND 
{b,k}O(notifyProductReceipt)_/F/_) AND {c,b}O(deliverProduct)_/F/_) AND 
{s,k}O(payShippingCosts)_/F/_ AND {s,c}O(sendProduct)_/F/_
\end{lstlisting}\\
		\hline
	\end{tabular}
	\caption{A contract decomposition}
	\label{fig:src_conflict}
\end{figure}


\section{RECALL Tool}\label{sec:implementation}

\begin{figure}[hbt]
	\centering
	\includegraphics[width=7cm]{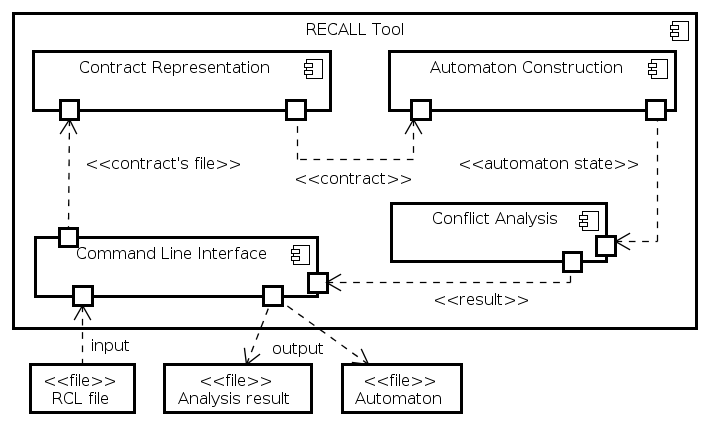}
	\caption{The RECALL architecture}
	\label{fig:modules}
\end{figure}

The conflict detection method proposed in Section~\ref{sec:method} has been implemented to allow automatic analysis for multi-party contracts specified on \RCL. 
The fundamental structures that composes the RECALL tool is presented in Figure~\ref{fig:modules}. 
The \emph{Command Line Interface} module implements the interaction between the tool and users, where settings for a contract analysis need to be provided to the tool.
The data information of a contract specified by the \RCL syntax are stated by the \emph{Contract Representation} module. 
The \emph{Automaton Construction} module constructs the automaton that models the contract described in \RCL. 
The \emph{Conflict Analysis} module aims at finding conflicts according to decompositions while the automaton is constructed. 

\subsection{Development}\label{subsec:implementation}

RECALL is mainly developed in Java~\cite{Java8}, but it is supported by several libraries such as \textit{ANTLR}~\cite{ANTLR}, which generates the grammar syntax, and \textit{Apache Commons CLI}~\cite{ApacheCLI}, which configures the user interface file. 
Some mathematical operations of set theory that are used by Automaton Construction and Conflict Analysis modules were optimized by \textit{Guava}~\cite{Guava} library. 
Further, the generated automaton of the analysis can also be graphically  yield due to the \textit{Graphviz}~\cite{Graphviz} library. 

The core method  of the construction automaton is shown at Figure~\ref{fig:src_constructAutomaton}. 
We highlight  some important fragments of this method. 
At line (4) we see that the Conflict Analysis module is invoked for every decomposition obtained by the automaton construction.
It allows to stop the verification process as soon as a conflict is found, avoiding unnecessary subsequent decompositions.
Nevertheless, the verification process may also continue up to the end of construction, when the yielded automaton represents the whole contract. 
All possible conflicts can then be revealed as seen at lines (7) to (9) of Figure~\ref{fig:src_constructAutomaton}. 
A contract is declared free of conflicts when all decompositions are conflict-free. 
Otherwise, the contract is deemed in conflict and a counter-example is displayed by the tool.
\begin{figure}[hbt]
	\centering
	\begin{tabular}{|p{\linewidth}|}
		\hline
        \vspace*{-1em}
\begin{lstlisting}[language=Java]
void constructAutomaton(State s) {
	Clause c1 = s.getClause();
	if (c1.getValue() == null) {
		if (searcher.hasConflict(s))
			automaton.setConflictFound(true);
		for (Set<RelativizedAction> a : generateActions(c1)) {
			if ((automaton.isConflictFound() && !config.isContinueOnConflict())
			|| s.getSituation() == StateSituation.conflicting)
				return;
			Clause c2 = decomposer.decompose((Clause) c1.clone(), a);
			State s1 = automaton.getStateByClause(c2);
			if (s1 != null) {
				automaton.getTransitions().add(new Transition(s, s1, a));
			} else {
				s1 = new State(c2);
				automaton.getStates().add(s1);
				automaton.getTransitions().add(new Transition(s, s1, a));
				constructAutomaton(s1);
				s1.getTrace().add(new Transition(s, s1, a));
			}
		}
	} else
		s.setSituation(
			c1.getValue()?StateSituation.satisfaction:StateSituation.violating);
}
\end{lstlisting}\\
		\hline
	\end{tabular}
	\caption{The automaton construction method}
	\label{fig:src_constructAutomaton}
\end{figure}
Decompositions are successively obtained by the \emph{Decomposer} class at line (10) according to the decomposition function $f$ (See Section~\ref{sec:method}). 
The \textit{decompose} method receives as parameters a clause that must be decomposed and a set of relativized actions and then returns a new clause following the rules defined in \RCL semantics. 
A new state is added to the automaton in order to represent the decomposed clause. 

The conflict detection algorithm is invoked for every decomposition of the automaton construction.  
This algorithm is implemented by the method \emph{hasConflict} and depicted at Figure~\ref{fig:src_hasConflict}. 
\begin{figure}[hbt]
	\centering
	\begin{tabular}{|c|}
		\hline
\begin{lstlisting}[language=Java]
boolean hasConflict(State state) {    
 Clause clause = state.getClause();
 Set<Set<DeonticTag>> delta = extractTags(clause);
 for (Set<DeonticTag> d1 : delta)
  for (Set<DeonticTag> d2 : Sets.difference(delta, Sets.newHashSet(d1)))
   for (DeonticTag d : d1) {
    Set<DeonticTag> conflictSet = generateConflictSet(d);
    Set<DeonticTag> inter = Sets.intersection(conflictSet, d2);
    if (!inter.isEmpty()) {
     state.setSituation(StateSituation.CONFLICTING);
     state.setConflictInformation(new ConflictInformation(d,inter,d2));
     return true;
    }
   }
 state.setSituation(StateSituation.CONFLICTFREE);
 return false;
}
\end{lstlisting}\\
		\hline
	\end{tabular}
	\caption{The conflict detection method}
	\label{fig:src_hasConflict}
\end{figure}
We see at line (3) that the \emph{extractTags} method implements the deontic extractor function $f_d$~\cite{rcl-sccc15}. 
This method receives a clause and returns sets of deontic tags according to  the evaluation of the current clause.
Every element of these sets of deontic tags, e.g., $d_i$, is checked against each element of the other sets, e.g., $d_j$, to finding potential conflicts.
Function $f\#$ returns any conflict between deontic operators related to $d$, for all $d \in d_i$, and any $d' \in d_j$, for all $d_j \neq d_i$, according to pre-defined conflicts. 
A conflict is raised by the tool if at least one element of $d_i$ is conflicting to an element of $d_j$.
In this case \emph{hasConflict} method, at line~(10), adds a flag at the state that contains the conflicting decompositions and stores the list of conflicting deontic tags, at line~(11), to be returned by the tool.

In the process of checking a multi-party contract, 
RECALL tool loads a text file with the contract's description in \RCL. 
The set of individuals $\mathcal{I}$ and the set of actions $\mathcal{A_B}$ are extracted to calculate the set of relativized actions $\mathcal{A}_r$ and also the set of concurrent actions ${\mathcal{A}_r^2}$. 
The sets of concurrent relativized actions are then ordered according to number of actions. 
We firstly consider the largest sets  to be processed in the process. 
The ordering strategy potentially increases the possibility of finding conflicts as soon as possible once a conflict becomes more likely to happen 
the bigger the number of concurrent actions.

The combinatorial computing of relativized actions grows exponentially as the number of actions and individuals also grows. 
But we notice that many of these combinatorial computation need not to be calculated since their representations do not reflect sceneries of the contract  under test. 
States and transitions of the automaton are constructed unnecessarily when theses sceneries are considered on the analysis. 
To illustrate this problem one considers the contract with $|\mathcal{I}|=4$ and $|\mathcal{A_B}|=3$. 
Hence the set of relativized actions $|\mathcal{A}_r|=48$ and the set of concurrent actions $|\mathcal{A}_r^2|=2^{48}-1$. 
We easily see that not only the ordering strategy over actions is important but also it is needed to overcome the combinatorial explosion problem.  
In any case we make clear that the nature of the problem remains combinatorial. 
We discuss aspects related to processing time and scalability in Section~\ref{sec:experiments}.  

We also apply a pruning strategy over the set of relativized actions when precessing clauses. 
We select on-the-fly only relativized actions related to the current decomposition on the process. 
Note that only actions of a current subcontract must be considered in the computation from that point onwards for corresponding decompositions. 
Irrelevant actions with respect to the current clause in a decomposition are then  discarded in subsequent evaluations. 
Decompositions over dynamic modalities need to take into account only trigger actions. 
Thus only one  particular state among those possible decompositions need to be computed, and the remaining clause is discarded in the current computation.  
Penalties of deontic modalities are also discarded in an evaluation since they take effect only when the obligation (or prohibition) is violated.

\subsection{Tool Practical Application}\label{subsec:application}

RECALL\footnote{More information and the source code is available on \url{http://recallcontracts.github.io}} has been conceived to analyze multi-party contracts specified in \RCL.
A multi-party contract must be described in a text file according to the \RCL syntax. 
Then the specification is submitted to the tool which, in turn, analyses the contract. 

The specification file is arranged in two parts: pre-defined conflicts and clauses.
Figure~\ref{fig:rcl_exemplo} presents a simple example of a contract specification file.
\begin{figure}[htb]
    \centering
    \begin{framed}
\begin{verbatim}
conflict {
    global { (a, b), (c, d) };
    relativized { (e, f), (e, a) };
};
[e]({j,k}O(f) ^ P(a) ^ {k}[a.b]({i,j}O(e&f)));
{j,i}F(c) _/{j}O(d)/_ ^ P(b) ^ {i,k}[a]({k}[b]({j,i}P(h))));
\end{verbatim}
    \end{framed}
    \caption{A simple example of a contract in \RCL.\label{fig:rcl_exemplo}}
\end{figure}
See that the pre-defined conflicts are described in the header of the file and it is optional once a contract may not have pre-defined conflicts. 
We note that global conflicts are given pairwise within the tag \emph{global}, and relativized conflicts  are defined pairwise within the tag \emph{relativized}.
Clauses of the contract are specified in sequel following the \RCL grammar. 

Relativization on deontic operators can be defined as follows:
\begin{enumerate}
    \item relativized, when there is only one individual between braces, e.g. \verb|{i}O(a)|, where individual $i$ is obliged to perform action $a$.
    \item directed, when there are two individuals related to the modality,  e.g. \verb|{i,j}O(a)|, where individual $i$ is obliged to perform action $a$ to individual $j$.   
    \item global, if the information is omitted, e.g. \verb|O(a)|, where all individuals of the contract are obliged to perform action $a$.
\end{enumerate}
Similarly we have relativizations over dynamic operators. 
Deontic modalities are represented by \verb|O|, \verb|P|, e \verb|F|, respectively, obligation, permission and prohibition operators.
Penalties are described between $\_/$ and $/\_$ after the associated clause. 

Once the contract has been specified we submit it to RECALL by setting the parameters of analysis.
Table~\ref{tab:cli} lists the running parameters of the tool. 
\begin{table}[hbt]
	\centering
	\begin{tabular}{|c|p{10cm}|}
		\hline
		\textbf{Parameter} & \textbf{Description}\\
		\hline
	$-c$ & the automaton is completely constructed and all conflicts can be found\\ 
		\hline
	$-g$ & the resulting automaton is exported to a DOT file\\ 
		\hline
	$-n$ & one considers all possible combinations of actions \\
		\hline
	$-v$ & the resulting analysis is displayed in verbose mode\\ 
		\hline
	$-h$ & the running options and some examples are displayed\\ 
		\hline				
	\end{tabular}
	\caption{Running parameters of RECALL tool\label{tab:cli}}
\end{table}
After running the tool, the resulting analysis says whether the contract under test is conflict-free. 
We obtain the output verdict by means of conflict detection traces (See Figure~\ref{fig:src_output})
and also  by a graphical representation of the constructed automaton. 

\subsection{Providing a Conflict-free Sales Contract}

The result of the verification analysis allows us to solve the raised conflict changing some deals in the original contract.
We first include a new clause where the bank now must notify the carrier with respect to the payment of the shipping costs performed by the seller.
Another amendment in the contract has been done specifically in the internal rule of the carrier.
Now the carrier takes into account the bank notification as a guarantee of payment for shipping costs.

We then rewrite the contract specification  following the proposed changes in order to avoid the conflict situation.
We change clauses (5) and (12) of the contract's description as follows:
\begin{framed}
{\scriptsize
\begin{enumerate}
    \item [] \textbf{[General Contract]}
    \setcounter{enumi}{4}
    \item \textbf{Bank} must notify the \textbf{Carrier} about the payment with respect to the shipping costs and after \textbf{Bank} attests the payment, the \textbf{Carrier} is obliged to deliver the product to the \textbf{Buyer}.
    \vspace*{2ex}
    \item [] \textbf{[Carrier Rules]}
    \setcounter{enumi}{11}
    \item \textbf{Carrier} is prohibited to deliver the product till \textbf{Bank} has notified  the Carrier that \textbf{Seller} has paid the shipping costs.
\end{enumerate}
* Note that the remaining clauses of the original contract is kept unchanged.
}
\end{framed}

The amendments of the new contract are presented in Figure~\ref{fig:src_studycase_free}.
The specific clause where the bank notifies the carrier with respect to the  shipping costs is described at line (2), which represents line (9) in the original contract.
The carrier considers the bank notification as guarantee to release the product delivery at line (4) in the new version corresponding to line (21) in the original contract.
Lines (1)-(8) and (10)-(20) of the original contract remain unchanged.
\begin{figure}[!htpb]
	\centering
	\begin{tabular}{|c|}
		\hline
\begin{lstlisting}[language=rcl]
...
(09)  {k,c}O(notifyShippingPayment) ^ {s,c}[sendProduct](
...
(21) {k,c}[(!notifyShippingPayment)*]({c,b}F(deliverProduct));
\end{lstlisting}\\
		\hline
	\end{tabular}
		\caption{The amended version of the sales contract}
		\label{fig:src_studycase_free}
\end{figure}

On the next step we resubmit the new version of sales contract to RECALL tool.
This time the resulting analysis has declared that the sales contract is conflict-free.
After these adjustments once the  bank is now obliged to notify the carrier about the payment of the shipping costs we guarantee that the contract runs to  completion safely without no loss to any party.
We remark that no other conflict has been detected in the contract.

For the sack of completeness we note that even the latest version of the sales contract cannot be captured by a set of bilateral contracts due to similar arguments that have been given beforehand over the oldest version.


\section{Practical Experiments}\label{sec:experiments}

We performed some experiments in order to verify the efficiency of our method and the tool scalability.
The experiments are classified according to the number of actions and individuals modeled in the contract specifications. 
We basically explore two main aspects on each group: \emph{execution time} and \emph{memory consumption}.
We separated the experiments into three groups, each group described in a corresponding subsection in the sequel.
In the first group we studied how the number of actions impacts in the verification process; in the second group  the number of individuals in the contracts were varied; and in the third we consider a high number of actions and individuals, simultaneously, in order to stress the tool up to its limit.

All experiments were performed using randomly generated contract specifications which were checked on a Core2duo running Linux with  8Gb of RAM memory.  
For each variation of a specific parameter, a total of 100 contract specifications were generated and checked.
We chose randomized experiments in order to avoid biases when treating each set of experiments.

\subsection{Varying the Number of Actions}

In this first group of experiments we varied the number of actions for each contract specification, with fixed number of individuals. 
The number of actions ranged from 8 to 15 and the number of individuals was fixed at 8. 
The execution time and the memory consumption is given by the average of 100 generated contracts for each class. 
For instance, the processing time takes at most 1.5 minutes for contracts with 8 individuals and 11 actions for 100 contract specifications, on average. 

\begin{figure}[!hbt]
\centering
\includegraphics[scale=0.5]{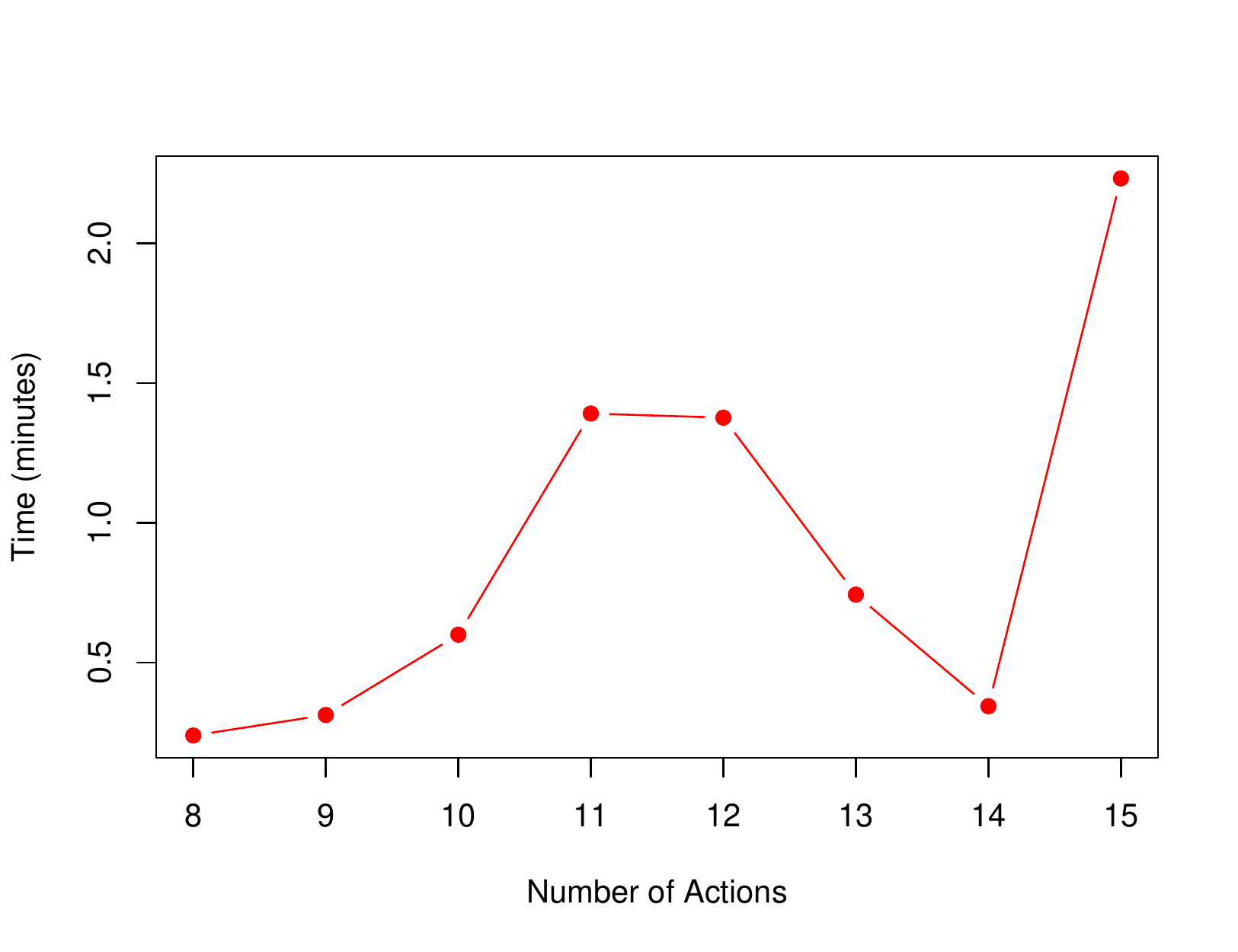}
\caption{Varying the number of actions: execution time}
\label{exp-actions-time}
\end{figure}

Figure~\ref{exp-actions-time} shows how the execution time of the verification process varied according to the number of actions in the contracts. 
We remark that some contract verifications  have not finished with a verdict due to the explosion problem, specially for contracts with a higher number of actions.
The  execution time average is  calculated based on the whole group of 100 contracts, seeing that for those unfinished experiments the resource consumption was much higher than the experiments with a verdict. 
For each class of contracts with 8 up to 15 actions, we obtained 2, 2, 3, 6, 5, 5, 2, and 13 unfinished checking processes of 100 performed experiments. 
We observe that the execution time, on average, does not take more than 2.5 minutes, in general, even for contracts with a high number of actions. 

\begin{figure}[!hbt]
\centering
\includegraphics[scale=0.5]{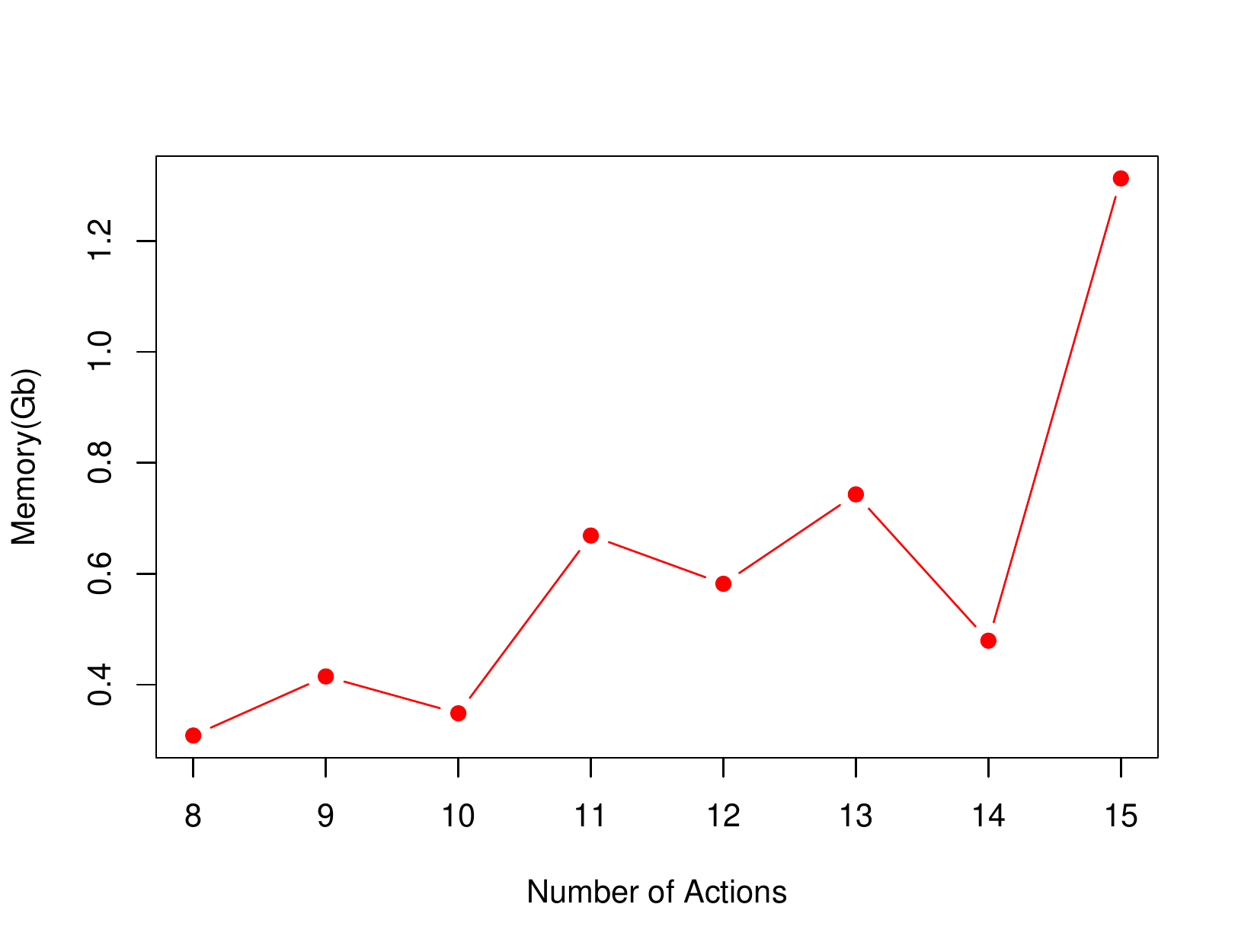}
\caption{Varying the number of actions: memory consumption}
\label{exp-actions-size}
\end{figure}

Similarly, we can observe the memory consumption for the same scenario in Figure~\ref{exp-actions-size}. 
We note that the memory consumption, on average, does not take more than around 1.4 Gb, in general. 
We also see that for the class of contracts with 8 individuals and 15 actions the memory consumption grows drastically compared to previous classes. 
We remark again that those unfinished checking processes are being calculated on the memory consumption average. 

\subsection{Varying the Number of Individuals}

In this group of experiments we varied the number of individuals in the contract specifications, with fixed number of actions. 
We ranged the number of individuals from 5 to 12 and the number of actions remained fixed at 10. 

\begin{figure}[!hbt]
\centering
\includegraphics[scale=0.5]{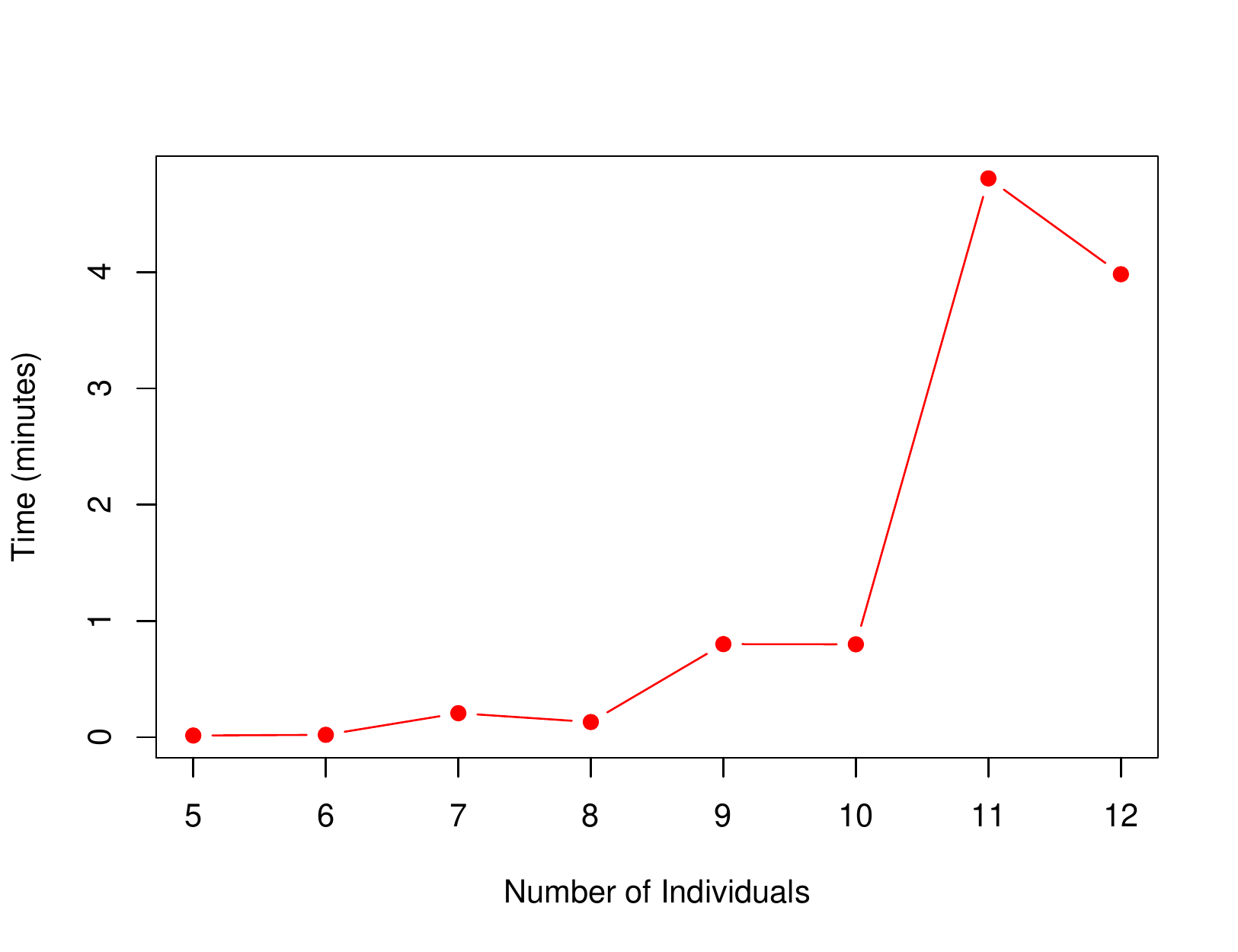}
\caption{Varying the number of individuals: execution time}
\label{exp-individuals-time}
\end{figure}

We show in Figure~\ref{exp-individuals-time} how the execution time of the verification process varied according to the number of individuals in the contracts. 
Again, we remark that the  execution time and memory average is calculated based on 100 contract specifications. 
In this scenario we notice that all experiments in groups with 5, and 6 individuals have finished with a verdict. 
Groups with 7 and 8 individuals had one unfinished checking process, while groups with 9, 10, 11, and 12 individuals had 6, 6, 16, and 20, respectively, unfinished processes. 
We observe that the execution time, on average, does not take more than 5 minutes, but we note that the execution time substantially grows from 11 individuals onwards. 

\begin{figure}[!hbt]
\centering
\includegraphics[scale=0.5]{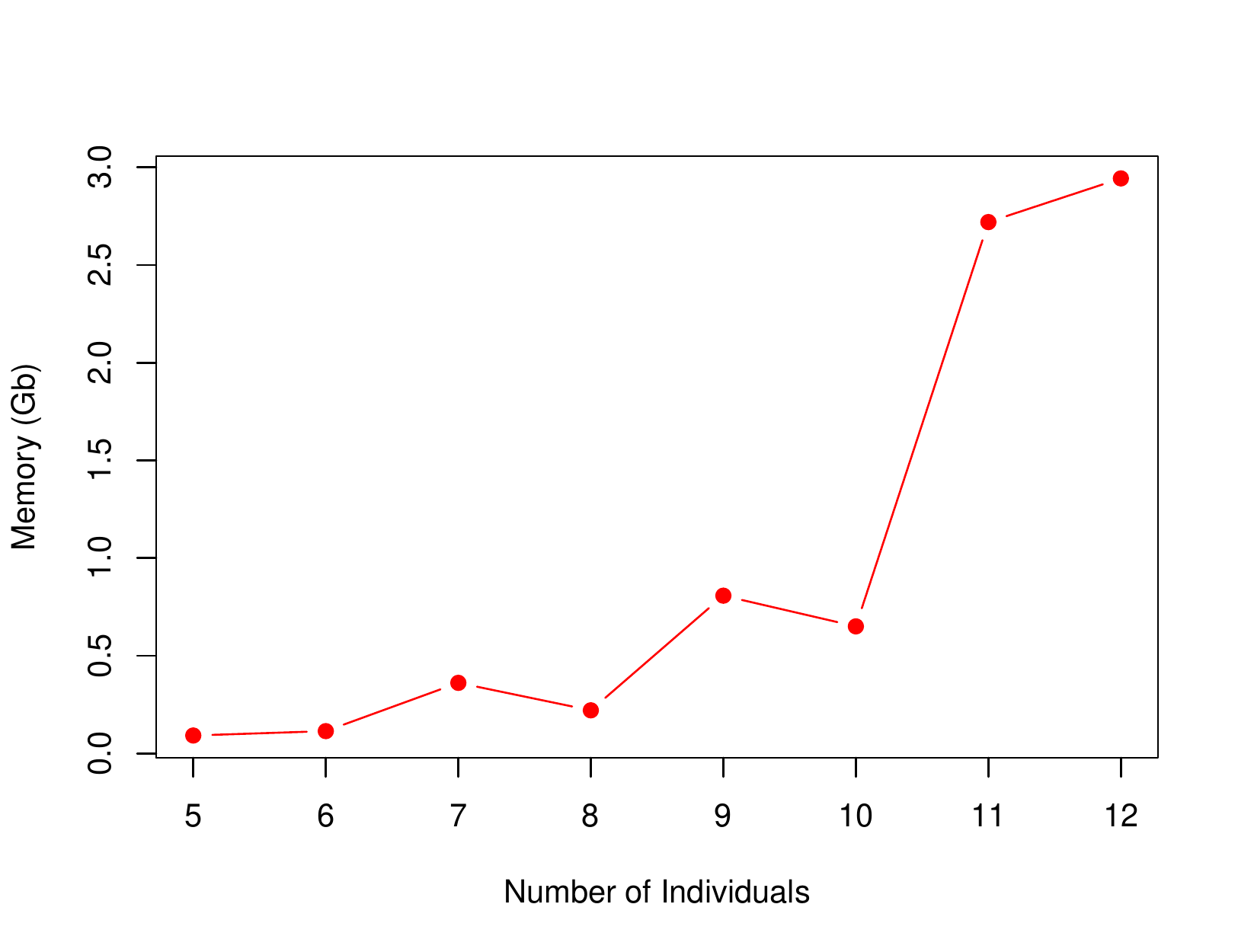}
\caption{Varying the number of individuals: memory consumption}
\label{exp-individuals-size}
\end{figure}

In Figure~\ref{exp-individuals-size} we observe the memory consumption for the same scenario. 
Similarly to the execution time the memory consumption drastically grows from experiments with 11 individuals onwards, more precisely, from 1 Gb to 3 Gb approximately. 
But, in general, the memory consumption does not take more than 3 Gb at maximum. 

\subsection{Stress testing}

In the last set of experiments  we have run contract specifications to stress the tool. 
We generated four groups of 100 contracts, but in this case we discarded unfinished checking runs. 

\begin{figure}[!hbt]
\centering
\includegraphics[scale=0.5]{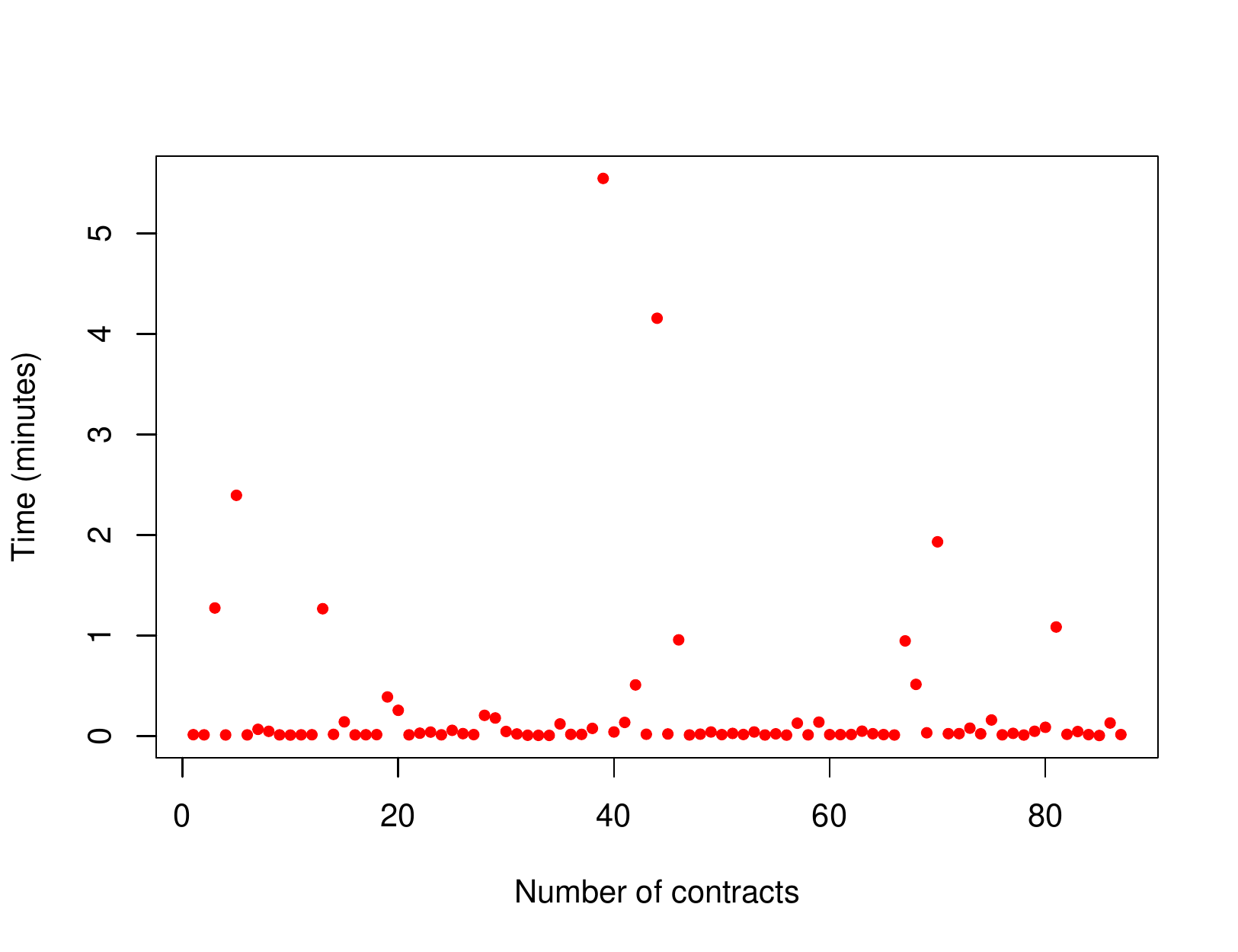}
\caption{Contracts with 8 individuals and 15 actions: execution time}
\label{exp-ind8-act15-time}
\end{figure}

\begin{figure}[!hbt]
\centering
\includegraphics[scale=0.5]{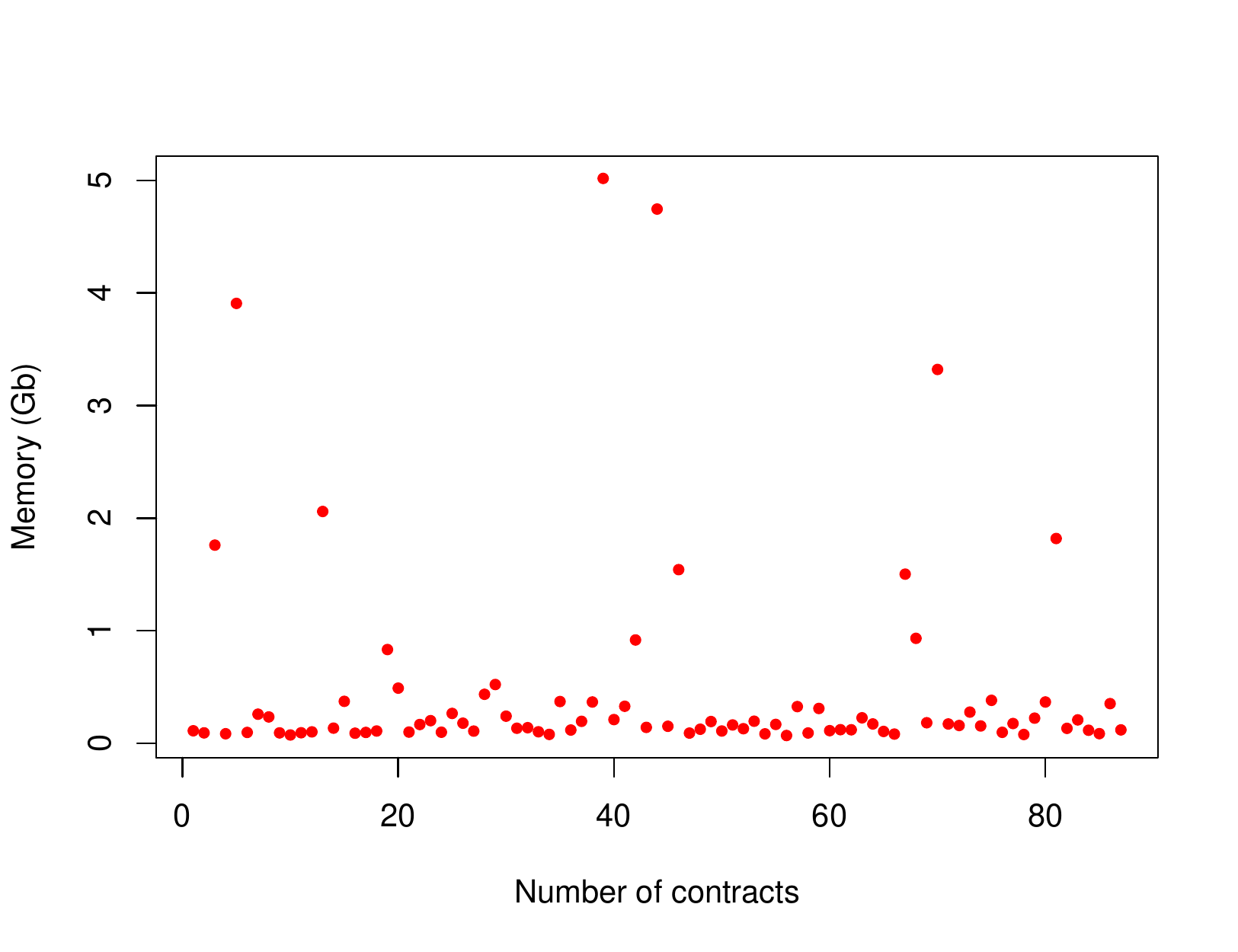}
\caption{Contracts with 8 individuals and 15 actions: memory consumption}
\label{exp-ind8-act15-size}
\end{figure}

In the first group we performed experiments where the contracts have  a fixed number of 8 individuals and a fixed number of 15 actions. 
The execution time on checking each contract is shown in Figure~\ref{exp-ind8-act15-time}. 
Here we have 13 contracts that do not run to completion.  
In contrast, we obtain 87 contract runs with a verdict, and most of them takes less than 30 seconds to be checked. 
We also observe the memory consumption for the same scenario in Figure~\ref{exp-ind8-act15-size}, and 
only 9 experiments from 87 finished runs take more than 1 Gb of  RAM memory. 

\begin{figure}[!hbt]
\centering
\includegraphics[scale=0.5]{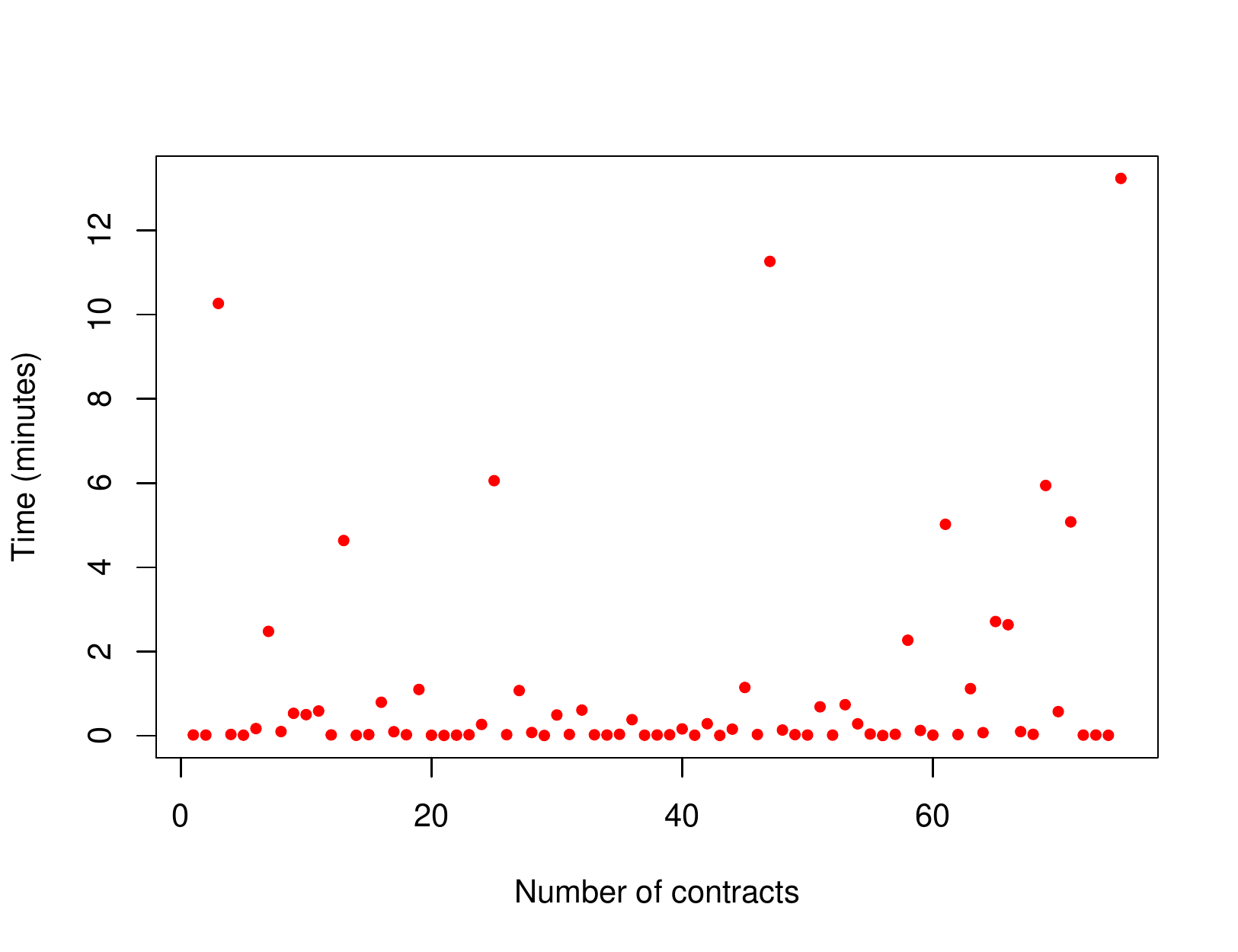}
\caption{Contracts with 8 individuals and 20 actions: execution time}
\label{exp-ind8-act20-time}
\end{figure}

Next we increase the number of actions from 15 to 20, while the number of individuals remains fixed at 8. 
We show in Figure~\ref{exp-ind8-act20-time} the execution time. 
We have 25 unfinished checking runs due to the memory overflow whereas 75 contracts have finished the process with a verdict, and only 10 \% (8 runs of 75 in total) has taken more than 3 minutes to be checked. 

\begin{figure}[!hbt]
\centering
\includegraphics[scale=0.5]{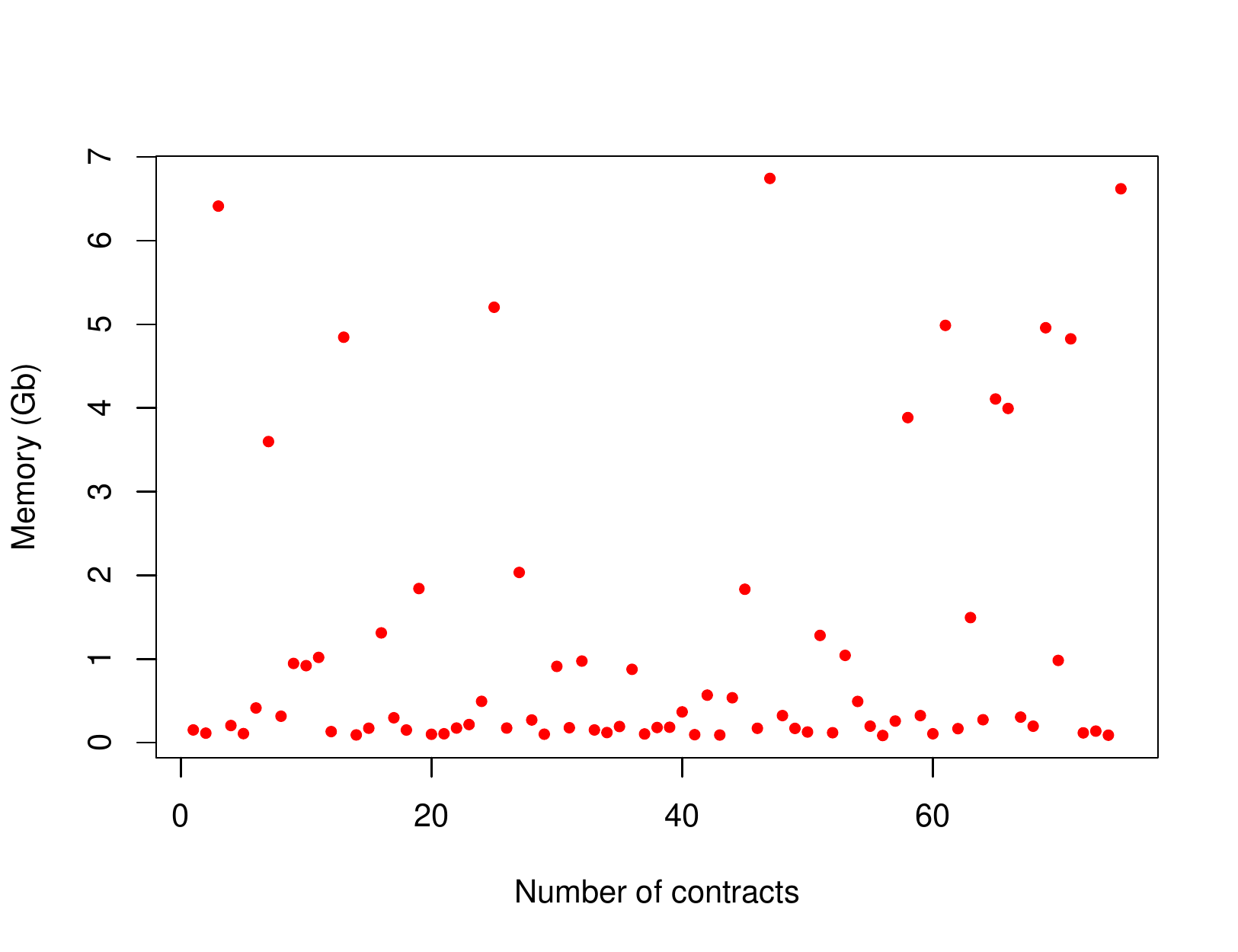}
\caption{Contracts with 8 individuals and 20 actions: memory consumption}
\label{exp-ind8-act20-size}
\end{figure}

We also observe that the memory consumption did not grow too much if compared to previous group, as seen in Figure~\ref{exp-ind8-act20-size}. 
Most of the experiments takes up to 2 Gb of memory at maximum, and only 16\%  of the experiments (12 runs of 75 in total)  have consumed between 2Gb and 7Gb of RAM memory. 

\begin{figure}[!hbt]
\centering
\includegraphics[scale=0.5]{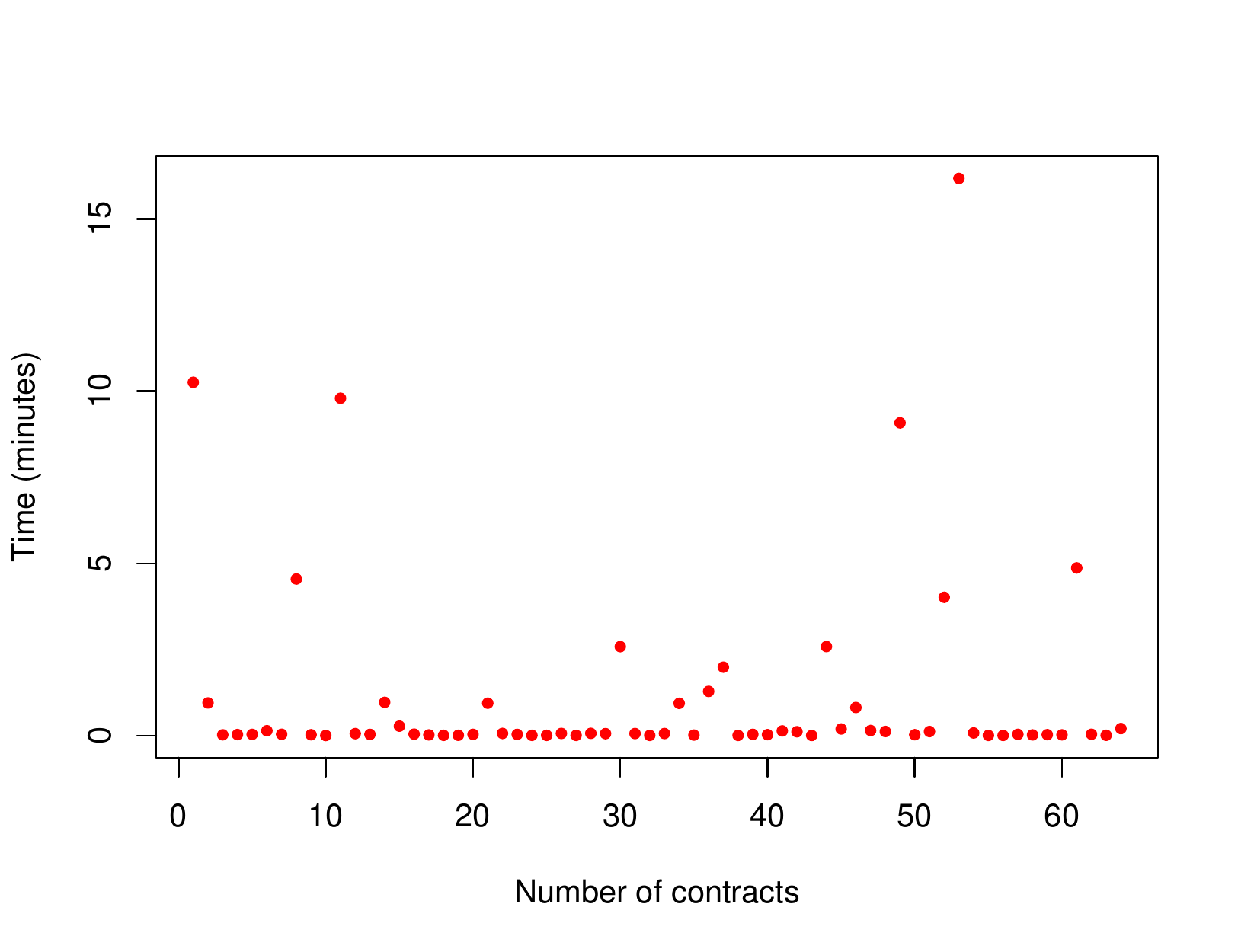}
\caption{Contracts with 15 individuals and 10 actions: execution time}
\label{exp-ind15-act10-time}
\end{figure}

In the third group we performed experiments where the contracts have  a fixed number of 15 individuals and a fixed number of 10 actions. 
That is, we increase the number of individuals, compared to the previous group, from 8 to 15, but we reduce the number of actions from 15 to 10 in total. 
Figure~\ref{exp-ind15-act10-time} shows the execution time on checking each contract. 
We reach 36 unfinished checking runs and 64 experiments that run to completion.  
We note that only 6\% (4 of 64 in total) of those finished  experiments take more than 5 minutes, and up to 17 minutes to be checked. 

\begin{figure}[!hbt]
\centering
\includegraphics[scale=0.5]{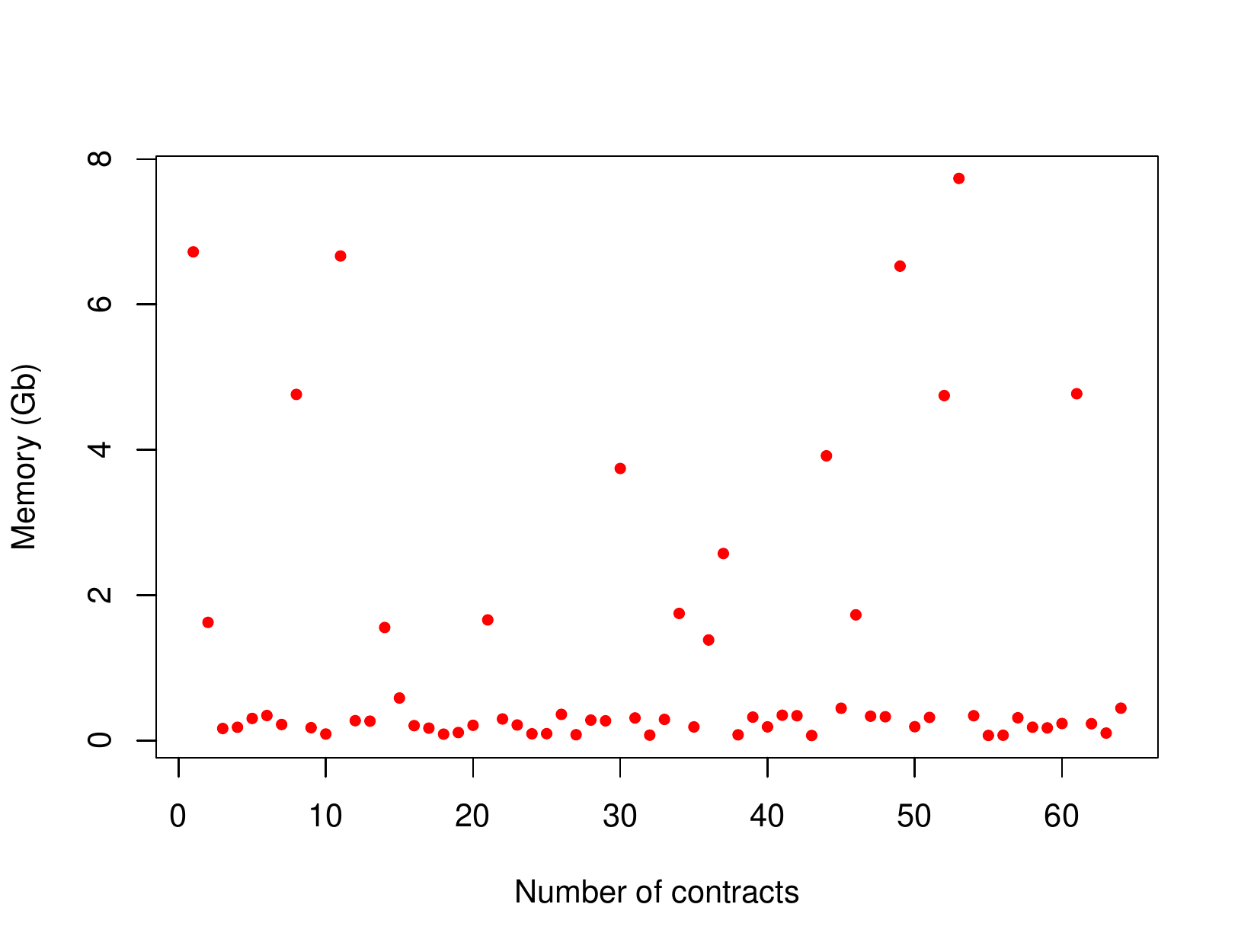}
\caption{Contracts with 15 individuals and 10 actions: memory consumption}
\label{exp-ind15-act10-size}
\end{figure}

By completeness Figure~\ref{exp-ind15-act10-size} presents the memory consumption for the same scenario. 
Most experiments also takes up to 2 Gb of memory at maximum, and only around 15\%  of the experiments (10 runs of 64 in total)  have consumed between 2Gb and 8Gb of  RAM memory. 

The resource consumption increases as the number of actions and individuals grows in the experiment. 
But also notice that the number of individuals has a more pronounced impact than the number of actions in the contracts.

\begin{figure}[!hbt]
\centering
\includegraphics[scale=0.5]{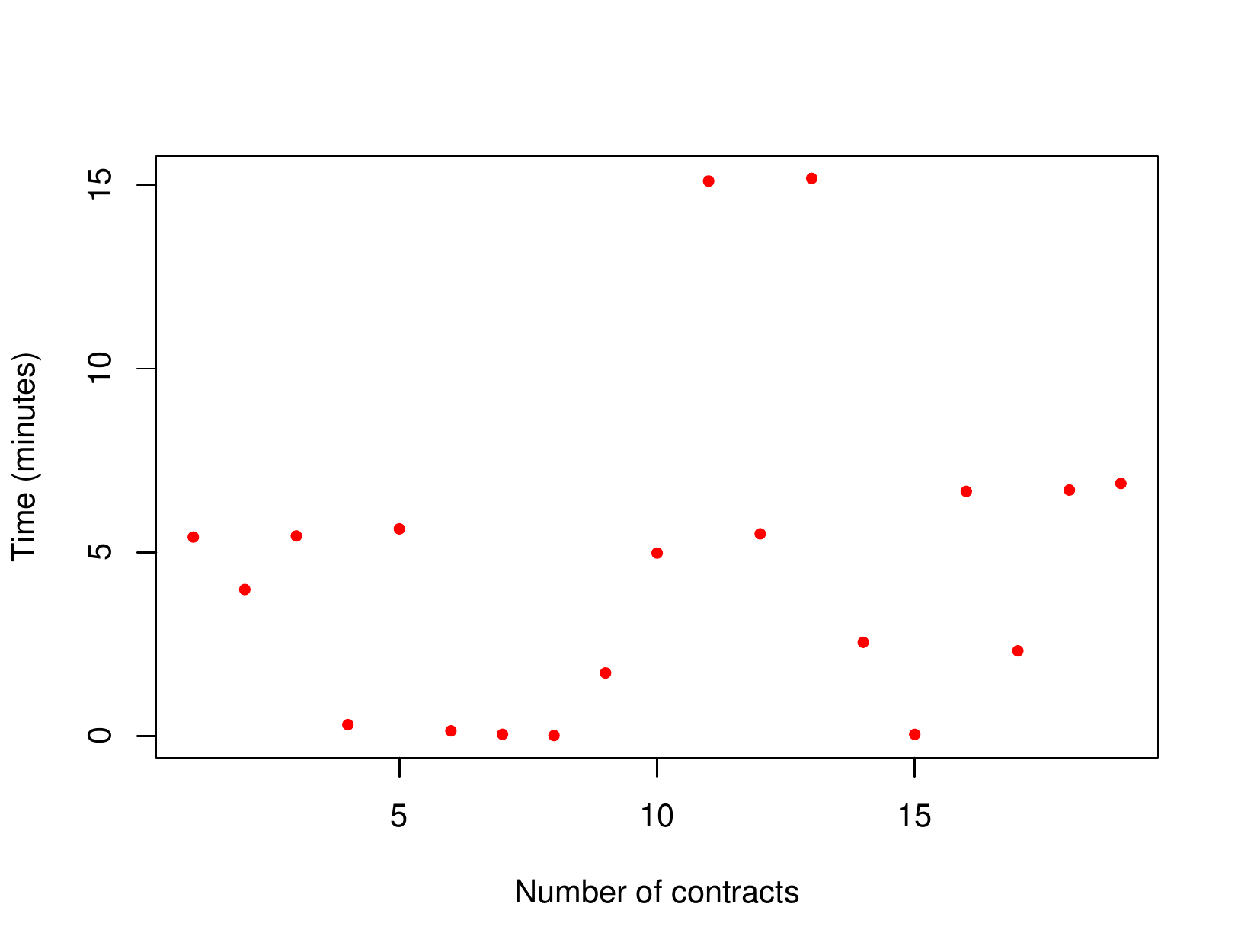}
\caption{Contracts with 18 individuals and 10 actions: execution time}
\label{exp-ind18-act10-time}
\end{figure}

In the last group we performed experiments where the contracts have  a fixed number of 18 individuals and a fixed number of 10 actions. 
That is, we increase the number of individuals, compared to the second group, from 8 to 18, and reduce the number of actions from 20 to 10. 
Figure~\ref{exp-ind18-act10-time}  shows the execution time. 
In this case we obtained at the rate of 81\% of unfinished checking runs while only 19 experiments have run to completion with the execution time up to 15 minutes. 

\begin{figure}[!hbt]
\centering
\includegraphics[scale=0.5]{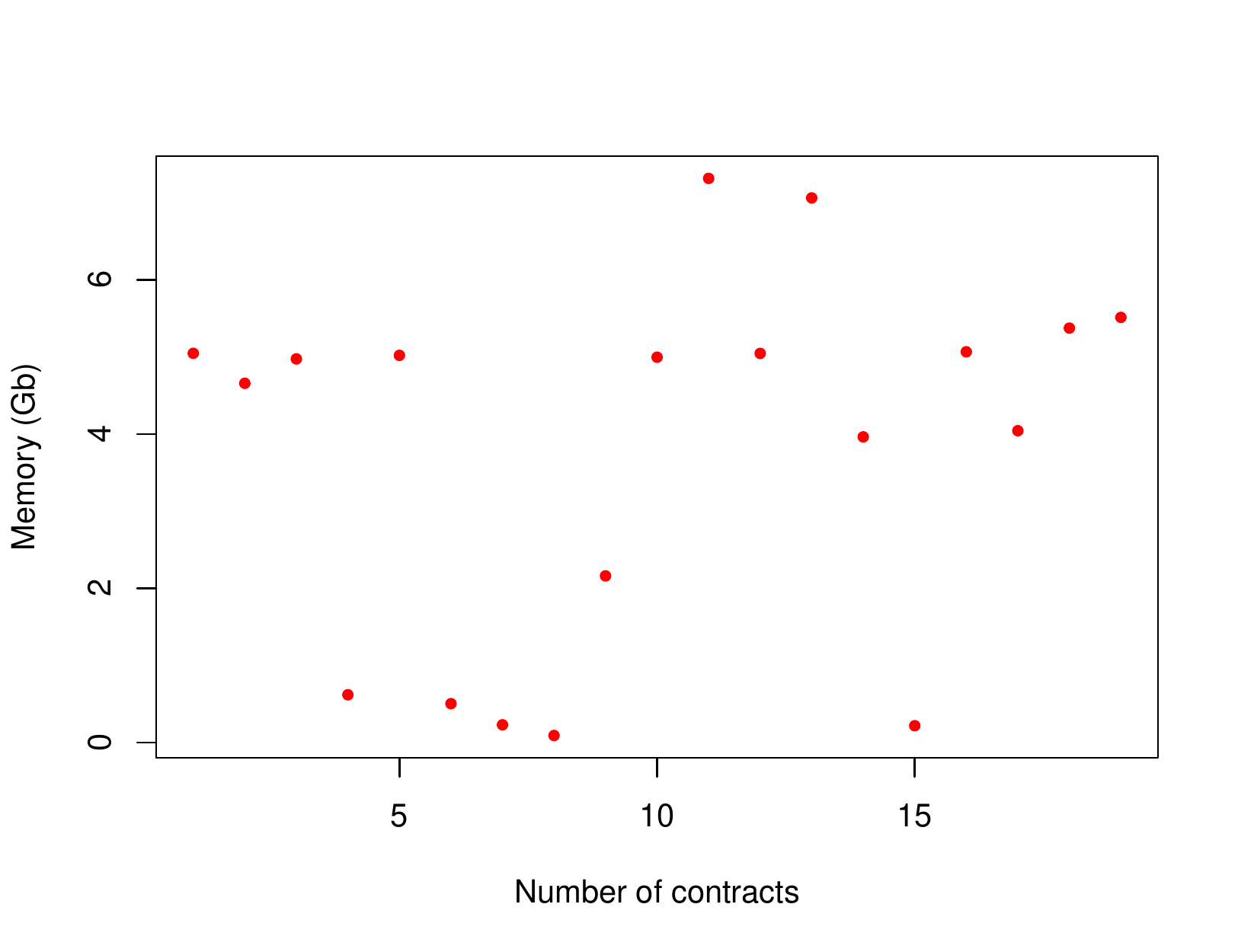}
\caption{Contracts with 18 individuals and 10 actions: memory consumption}
\label{exp-ind18-act10-size}
\end{figure}

Regarding the memory consumption we see in the Figure~\ref{exp-ind18-act10-size} that only 5 contracts were checked with less than 2Gb of RAM memory, and the remaining takes between 2Gb and 8Gb of RAM memory. 

We noticed a more prominently resource consumption when the number of individuals grows in the experiment. 
 In this last scenario, with 18 individuals and 10 actions, we observe higher ratios of execution time and memory consumption when the number of individuals and actions are inversely proportional compared to the second group where the contracts have 8 individuals and 20 actions. 
The result reinforces the effect regarding the growth of individuals more than the number of actions in the contract specifications. 

\subsection{Threats to Validity}

All contract specifications were randomly generated in our experiments.
Even though the generation procedure is unbiased, we cannot able to make claims about the similarity between these generated contract specifications and real contracts which may appear in practice. 
To overcome this threat, in future works, more experiments could be replicated using real contract specifications as made with the sales contract. 
Another threat is related to the algorithm  that randomly generate the contract specifications. 
It may somehow bring forth rarer and special cases where the RECALL tool does not deal with very well, \emph{e.g.} which logical operators have been randomly chosen within the clauses. 
We tackled this threat by using a high number of contracts for each group of experiments, reducing the influence of this factor on the results.


\section{Conclusion} \label{sec:conclusion}

This work proposed a method for checking multi-party contracts that can be formally modeled by an extended contract language. 
We defined the syntax and the semantics for t
The extended contract language, named  \RCL, was defined to enclose relativizations in the classical contract language. 
We developed the proposal to automatically check multi-party contracts specified by \RCL. 

A real-world contract, characterized by multi-party aspects, was also specified and submitted to our tool as a case study of an electronic commerce problem. 
The tool was able to detect an important problem in the original specification rising a conflicting relationship between their participating parties. 
We could fix the contract specification by means the resulting analysis and in a new checking run with our tool declared a 
conflict-free verdict over the reviewed version for the sales contract. 

We also performed practical experiments for different scenarios in order to evaluate the  RECALL tool. 
We assay aspects related to scalability  and efficiency of our tool, exploring complementary parameters, i.e. number of actions and number of individuals that are restrained on the contract. 
The experimental results indicated a good performance when we have a balance between the number of individuals and the number of actions summed up of around 20.
The analyses also revealed that the number of individuals has greater influence than the number of actions on this efficiency.

We leave for future work a  graphical user interface to ease the contract modeling and the process of analysis. 
We also expect to inspire other works to improve the algorithms proposed in this work.


\begin{thebibliography}{10}

\bibitem{Angelov01b2becontract}
S.~Angelov and P.~Grefen.
\newblock B2b econtract handling - a survey of projects, papers and standards.
\newblock Technical report, University of Twente, The Netherlands, 2001.

\bibitem{Daskalopulu2001}
Aspassia Daskalopulu.
\newblock Model checking contractual protocols.
\newblock {\em CoRR}, cs.SE/0106009, 2001.

\bibitem{RT-DC-15-01}
Wellington~A. {Della Mura} and Adilson~L. Bonif{\'{a}}cio.
\newblock {A conflict detection approach for multi-party contracts}.
\newblock Technical Report DC-15-01, Department of Computing, University of
  Londrina, August 2015.
\newblock In English, 14 pages.

\bibitem{rcl-sccc15}
Wellington~Aparecido {Della Mura} and Adilson~Luiz Bonif{\'{a}}cio.
\newblock Devising a conflict detection method for multi-party contracts.
\newblock In {\em 34th International Conference of the Chilean Computer Science
  Society, {SCCC} 2015, Santiago, Chile, November 9-13, 2015}, pages 1--6.
  {IEEE}, 2015.

\bibitem{Fenech09Automatic}
Stephen Fenech, GordonJ. Pace, and Gerardo Schneider.
\newblock Automatic conflict detection on contracts.
\newblock In Martin Leucker and Carroll Morgan, editors, {\em Theoretical
  Aspects of Computing - ICTAC 2009}, volume 5684 of {\em Lecture Notes in
  Computer Science}, pages 200--214. Springer Berlin Heidelberg, 2009.

\bibitem{Graphviz}
Emden Gansner.
\newblock Graphviz 2.38, 2016.

\bibitem{Guava}
Google.
\newblock Guava 1.8, 2015.

\bibitem{Harel84dynamiclogic}
David Harel, Dexter Kozen, and Jerzy Tiuryn.
\newblock Dynamic logic.
\newblock In {\em Handbook of Philosophical Logic}, pages 497--604. MIT Press,
  1984.

\bibitem{Haugen02_multy}
B~Haugen.
\newblock Multi-party electronic business transactions.
\newblock {\em
  http://www.supplychainlinks.com/MultiPartyBusinessTransactions.PDF}, 2002.

\bibitem{Herrestad1995Deontic}
Henning Herrestad and Christen Krogh.
\newblock Obligations directed from bearers to counterparts.
\newblock {\em ICAIL '95 Proceedings of the 5th international conference on
  Artificial intelligence and law}, pages 453--522, 1995.

\bibitem{Hilpinen2001Deontic}
Risto Hilpinen.
\newblock Deontic logic.
\newblock In Lou Goble, editor, {\em The blackwell guide to philosophical
  logic}, chapter~8. Blackwell, 2001.

\bibitem{Kyas08runtimemonitoring}
Marcel Kyas, Cristian Prisacariu, and Gerardo Schneider.
\newblock Runtime monitoring of electronic contracts.
\newblock In {\em In ATVA08, LNCS}. Springer-Verlag, 2008.

\bibitem{Meyer1987ADifferentApproach}
J.~J.~Ch Meyer.
\newblock A different approach to deontic logic: Deontic logic viewed as a
  variant of dynamic logic.
\newblock {\em Notre Dame Journal of Formal Logic}, 29(1):109--136, 1987.

\bibitem{murata89}
T.~Murata.
\newblock Petri nets: Properties, analysis and applications.
\newblock {\em Proceedings of the IEEE}, 77(4):541--580, Apr 1989.

\bibitem{Java8}
Oracle.
\newblock Java se development kit 8, 2015.

\bibitem{ANTLR}
Terence Parr.
\newblock Antlr 4.5, 2015.

\bibitem{DBLP:conf/ifip/Petri62}
C.~A. Petri.
\newblock Fundamentals of a theory of asynchronous information flow.
\newblock In {\em {IFIP} Congress}, pages 386--390, 1962.

\bibitem{Prisacariu07aformal}
Cristian Prisacariu and Gerardo Schneider.
\newblock A formal language for electronic contracts.
\newblock In {\em In FMOODS’07, volume 4468 of LNCS}, pages 174--189.
  Springer, 2007.

\bibitem{Prisacariu09AnAction}
Cristian Prisacariu and Gerardo Schneider.
\newblock Cl: An action-based logic for reasoning about contracts.
\newblock In Hiroakira Ono, Makoto Kanazawa, and Ruy Queiroz, editors, {\em
  Logic, Language, Information and Computation}, volume 5514 of {\em Lecture
  Notes in Computer Science}, pages 335--349. Springer Berlin Heidelberg, 2009.

\bibitem{Prisacariu12_a_dynamic}
Cristian Prisacariu and Gerardo Schneider.
\newblock A dynamic deontic logic for complex contracts.
\newblock {\em The Journal of Logic and Algebraic Programming}, 81(4):458 --
  490, 2012.

\bibitem{Reisig:1985:PNI:3405}
Wolfgang Reisig.
\newblock {\em Petri Nets: An Introduction}.
\newblock Springer-Verlag New York, Inc., New York, NY, USA, 1985.

\bibitem{ApacheCLI}
{The Apache Software Foundation}.
\newblock Apache commons cli 1.3.1, 2015.

\bibitem{Xu04amulti-party}
Lai Xu.
\newblock A multi-party contract model.
\newblock {\em ACM SIGecom Exchanges}, 5, 2004.

\bibitem{Xu04_monitoring_thesys}
Lai Xu.
\newblock {\em Monitoring Multi-party contracts for e-business}.
\newblock Tese, Faculty of Economics and Business Administration of Tilburg
  University, 2004b.

\end{thebibliography}
\end{document}